\documentclass[12pt]{article}
\usepackage{amsmath, amssymb, amsthm, amsfonts, latexsym, graphicx, color}
\usepackage{authblk}
\usepackage{hyperref}
\usepackage{adjustbox}
\usepackage{latexsym, stmaryrd}
\usepackage{enumitem}

\newlength{\actualtopmargin}
\newlength{\actualsidemargin}
\newtheorem{lemma}{Lemma}
\newtheorem{claim}{Claim}

\newtheorem{fact}{Fact}

\newtheorem{definition}{Definition}

\newcommand{\ket}[1]{\left|#1\right\rangle} 
\newcommand{\kets}[1]{|#1\rangle} 
\newcommand{\bra}[1]{\left\langle #1\right|} 
\newcommand{\bras}[1]{\langle #1|} 
\newcommand{\braket}[2]{\langle #1 | #2 \rangle} 

\newcommand{\blnk}{ 
\kern-1bp
\setlength{\unitlength}{10bp}
\begin{picture}(1,1)
\put(0.15,0.05){$\bigcirc$}
\end{picture}
\kern+3bp
}
\newcommand{\lmove}{ 
\setlength{\unitlength}{10bp}
\kern-1bp
\begin{picture}(1,1)
\put(0.27,0.05){$\shortleftarrow$}
\put(0.15,0.05){$\bigcirc$}
\end{picture}
\kern+3bp
}
\newcommand{\turn}{ 
\setlength{\unitlength}{10bp}
\kern-1bp
\begin{picture}(1,1)
\put(0.27,0.05){$\circlearrowleft$}
\put(0.15,0.05){$\bigcirc$}
\end{picture}
\kern+3bp
}
\newcommand{\insi}{ 
\setlength{\unitlength}{10bp}
\kern-1bp
\begin{picture}(1,1)
\put(0.4,0.05){$\circ$}
\put(0.15,0.05){$\bigcirc$}
\end{picture}
\kern+3bp
}
\newcommand{\dead}{ 
\kern-1bp
\setlength{\unitlength}{10bp}
\begin{picture}(1,1)
\put(0.27,0.05){$\times$}
\put(0.15,0.05){$\bigcirc$}
\end{picture}
\kern+3bp
}
\newcommand{\qubit}{ 
\setlength{\unitlength}{10bp}
\kern+1bp
\begin{picture}(1,1)
\linethickness{1bp}
\put(0,-0.2){\line(0,1){1}}
\put(0,-0.2){\line(1,0){1}}
\put(1,0.8){\line(0,-1){1}}
\put(1,0.8){\line(-1,0){1}}
\end{picture}
\kern+1bp
}
\newcommand{\gate}{ 
\setlength{\unitlength}{10bp}
\kern+1bp
\begin{picture}(1,1)
\put(0.1,0){$\blacktriangleright$}
\linethickness{1bp}
\put(0,-0.2){\line(0,1){1}}
\put(0,-0.2){\line(1,0){1}}
\put(1,0.8){\line(0,-1){1}}
\put(1,0.8){\line(-1,0){1}}
\end{picture}
\kern+1bp
}
\newcommand{\rmove}{ 
\setlength{\unitlength}{10bp}
\kern+1bp
\begin{picture}(1,1)
\put(0.1,0){$\vartriangleright$}
\linethickness{1bp}
\put(0,-0.2){\line(0,1){1}}
\put(0,-0.2){\line(1,0){1}}
\put(1,0.8){\line(0,-1){1}}
\put(1,0.8){\line(-1,0){1}}
\end{picture}
\kern+1bp
}
\newcommand{\abet}{ 
\kern+1bp
\setlength{\unitlength}{10bp}
\begin{picture}(1,1)
\put(0.15,-0.05){$\beta$}
\linethickness{1bp}
\put(0,-0.2){\line(0,1){1}}
\put(0,-0.2){\line(1,0){1}}
\put(1,0.8){\line(0,-1){1}}
\put(1,0.8){\line(-1,0){1}}
\end{picture}
\kern+1bp
}
\newcommand{\alpa}{ 
\kern+1bp
\setlength{\unitlength}{10bp}
\begin{picture}(1,1)
\put(0.15,0.05){$\alpha$}
\linethickness{1bp}
\put(0,-0.2){\line(0,1){1}}
\put(0,-0.2){\line(1,0){1}}
\put(1,0.8){\line(0,-1){1}}
\put(1,0.8){\line(-1,0){1}}
\end{picture}
\kern+1bp
}
\newcommand{\lqubit}{ 
\kern+1bp
\setlength{\unitlength}{10bp}
\begin{picture}(1,1)
\put(0.15,0.05){$\scriptstyle{L}$}
\linethickness{1bp}
\put(0,-0.2){\line(0,1){1}}
\put(0,-0.2){\line(1,0){1}}
\put(1,0.8){\line(0,-1){1}}
\put(1,0.8){\line(-1,0){1}}
\end{picture}
\kern+1bp
}
\newcommand{\rqubit}{ 
\kern+1bp
\setlength{\unitlength}{10bp}
\begin{picture}(1,1)
\put(0.15,0.05){$\scriptstyle{R}$}
\linethickness{1bp}
\put(0,-0.2){\line(0,1){1}}
\put(0,-0.2){\line(1,0){1}}
\put(1,0.8){\line(0,-1){1}}
\put(1,0.8){\line(-1,0){1}}
\end{picture}
\kern+1bp
}
\newcommand{\parity}{ 
\setlength{\unitlength}{10bp}
\begin{picture}(0.3,1)
\put(0.15,-0.3){\line(0,1){1.2}}
\end{picture}
}
\newcommand{\bdry}{ 
\setlength{\unitlength}{10bp}
\begin{picture}(0.3,1)
\put(0.07,-0.3){\line(0,1){1.2}}
\put(0.23,-0.3){\line(0,1){1.2}}
\end{picture}
}
\newcommand{\bothp}{ 
\setlength{\unitlength}{10bp}
\begin{picture}(0.3,1)
\put(0.07,0){\line(0,1){0.3}}
\put(0.23,0){\line(0,1){0.3}}
\put(0.07,0.6){\line(0,1){0.3}}
\put(0.23,0.6){\line(0,1){0.3}}
\put(0.15,0.3){\line(0,1){0.3}}
\put(0.15,-0.3){\line(0,1){0.3}}
\end{picture}
}

\newcommand{\ok}{$\checkmark$} 
\newcommand{\xx}{-----}
\newcommand{\prodd}{$\,\cdot\, \parity \,\cdot\,$}
\newcommand{\preve}{$\parity \,\cdot\, \,\cdot\,\parity$}
\newcommand{\prbleft}{$\bdry \,\cdot\, \,\cdot\, \parity$}
\newcommand{\prbcent}{$\,\cdot\, \bdry \,\cdot\,$}
\newcommand{\prbrigh}{$\parity \,\cdot\, \,\cdot\,\bdry$}

\newcommand{\ii}{\mathbb{I}} 
\newcommand{\ppp}{\phantom{\parity}} 

\newcommand{\cH}{\mathcal{H}} 

\newcommand{\mgdots}{\cdots}

\begin{document}

\title{\Large
  \textbf{
     The Local Hamiltonian problem on a line\\with eight states is QMA-complete
  }
}

\author{SEAN HALLGREN\footnote{Partially supported by National Science Foundation grant CCF-0747274 and by the National Security Agency (NSA) under Army Research Office (ARO) contract number W911NF-08-1-0298.}}
\affil{{\small Department of Computer Science and Engineering, The Pennsylvania State University}}
\author{DANIEL NAGAJ\footnote{Partially supported by the Slovak Research and Development Agency grant COQI APVV-0646-10, EU project Q-ESSENCE 2010-248095, EU Commission ERC Starting Grant QUERG No. 239937. and by the Austrian FWF SFB grants FoQuS and ViCoM}}
\affil{{\small Faculty for Physics, University of Vienna,  and\\Institute of Physics, Slovak Academy of Sciences}}
\author{SANDEEP NARAYANASWAMI}
\affil{{\small Department of Physics, The Pennsylvania State University}}

\maketitle
\vspace{-5mm}

\begin{abstract}
The Local Hamiltonian problem is the problem of estimating the least
  eigenvalue of a local Hamiltonian, and is complete for the class
  QMA. The 1D problem on a chain of qubits has heuristics which work
  well, while the 13-state qudit case has been shown to be
  QMA-complete.  We show that this problem remains QMA-complete when
  the dimensionality of the qudits is brought down to 8.
\end{abstract}


\section{Introduction}
The Local Hamiltonian problem -- estimating the ground state energy of a
local Hamiltonian -- is a natural problem in physics, and belongs to
the complexity class QMA. QMA is the quantum analogue of NP. Languages
in QMA have a quantum verifier: a polynomial-time quantum algorithm
that takes (poly-sized) quantum states as witnesses.  

In quantum mechanics, the Hamiltonian of a system is the Hermitian
operator corresponding to the energy of the system: its eigenvalues
are the set of energies that a system can be measured to have. It also
determines the time-evolution of the system and defines the
interactions between its subsystems. The least eigenvalue (ground state
energy) and the corresponding eigenvector (the ground state) are key
to understanding the properties of a quantum system. Hamiltonians in
nature are usually local, in that they can be written as a sum $H =
\sum_{i = 1}^{M} H_{i}$ where each term $H_i$ acts only on a small
(constant) number of subsystems. Estimating the ground state energy of such
an $H$ is therefore a very fundamental question. 

The input to the $k$-Local Hamiltonian problem is a set of Hermitian
matrices $\{ H_i \}$, each $H_i$ acting on a set of $k$ qubits (out of
a total of $n$), and the problem is to estimate the lowest eigenvalue of the sum $H = \sum_{i = 1}^{M}
H_{i}$. Note that even though each $H_i$ acts nontrivially on a constant number
$k$ of qubits and is constant-dimensional, $H$ itself acts on the
whole space of $n$ qubits and is therefore exponential in size. The
Local Hamiltonian problem can be thought of as a generalization of
SAT~\cite{AN02}. In particular, MAX2SAT is a special case of the
2-Local Hamiltonian problem. Therefore, 2-Local Hamiltonian is NP-hard.

The 5-Local Hamiltonian problem was the first to be shown to be
QMA-complete, in~\cite{KSV02}. It is also a very natural complete
problem, given that it is a generalization of SAT. Moreover, physicists
have worked on similar problems, developing a number of heuristic
tools for approximating ground states and ground state energies. However,
the Hamiltonian constructed in~\cite{KSV02} does not have any constrain
the spatial arrangement of the qubits, making it unrealistic. 
In physical (e.g. spin) systems, the Hamiltonians are often spatially
local: the interacting systems (qubits or qudits) may be arranged on a
grid, or the interactions are (at least approximately)
short-ranged (e.g. nearest-neighbor). Simulation of local Hamiltonians on one- or
two-dimensional grids is an important problem in physics, and  it is
natural to try to understand the complexity in the different cases
obtained by changing the locality and the dimensionality of the
qudits. Since it is also much easier to realize and manipulate lower-dimensional qudits in the
lab, these cases are particularly important.  

There have been improvements on this result and its unfrustrated
variant Quantum $k$-SAT~\cite{Bra06,ER08,Nag10}. The locality was
brought down to 3~\cite{KR03} and then to 2~\cite{KKR06}. The 2-local
problem remains QMA-complete when the Hamiltonians are restricted to
be nearest-neighbor interactions on a 2D grid~\cite{OT08}. The 1D case
was not expected to be so hard: its classical counterpart, the 1D
constraint satisfaction problem, has efficient algorithms. Moreover,
there are good heuristic methods that are effective on many instances
of the problem. Therefore, it was a somewhat surprising result
when~\cite{AGIK09} showed a hardness proof for the Local Hamiltonian
problem on a chain (with nearest-neighbor interactions) of
13-state\footnote{The paper states hardness for $d=12$. However, there
  are two illegal configurations that are not penalized: $
  \dead\dead\lmove\blnk\blnk\ $ and $\dead\dead\turn\blnk\blnk\ $ turn
  into each other under the action of the Hamiltonian. The
  superposition of these two configurations forms a zero energy state of
  the Hamiltonian, which means that the Hamiltonian no longer has the
  promised $\frac{1}{poly}$ gap. This can be fixed by adding a 13th
  state, as discussed in footnote 4 in \cite{AGIK09}.} qudits. In this
paper, we bring the number of states down from 13 to 8. For a recent review of QMA-complete problems, see \cite{BOOKATZ}.

The hardness of the 1D problem (with nearest-neighbor interactions only) 
for the cases with 2-7 state qudits remains an interesting open question. It is not clear if the
QMA-completeness result will continue to hold as we further decrease the
dimensionality of the qudits down towards 2. It may happen that below a
particular dimensionality, we could find that the problem has an
efficient quantum or classical algorithm, e.g. if the ground state entanglement could be shown to be low. 
Recently, an interesting qutrit chain with a unique unfrustrated ground state with lots of entanglement 
was analyzed in \cite{BCMNS12}. 
Finally, these QMA-completeness results also bear a close relationship to adiabatic
quantum computing: the computation models in these results, and the
Hamiltonians that check these computations, can be used to perform
universal adiabatic quantum computing. It will be interesting to see
if restricted local Hamiltonian systems (e.g.,  low-dimensional qudits
on a line) that most likely do not encode 
a QMA-complete problem can still be used to perform universal adiabatic QC.  
  
To show QMA completeness for our version of Local Hamiltonian, we
reduce an arbitrary QMA language $L$ to a Local Hamiltonian in 1D with
$d=8$ particles, outputting a Hamiltonian that either has a ground state
energy below some value $a$, or whether this energy is at least $1/poly$ larger than $a$. 
We base our proof on three ideas. 

First, we use Kitaev's Hamiltonian~\cite{KSV02}: 
a Hamiltonian that has as its
ground state a \emph{history state} of the verification circuit\footnote{The subscript 
$x$ in the verifier circuit $V_x$ stands for the instance $x$ of the problem.} 
$V_x$ for the language $L$. A history state is a state of the form $\sum_{t} \vert \phi_{t}
\rangle \vert t \rangle$, where $\vert \phi_{t} \rangle$ is the state
after applying the first $t$ gates of $V_{x}$ to
$|\phi_0\rangle$. Kitaev's Hamiltonian induces forward and backward
transitions between consecutive time-steps, i.e., $\ket{\phi_t}
\ket{t} \longleftrightarrow \ket{\phi_{t+1}} \ket{t+1} $.  In
addition, the Hamiltonian serves to ensure that no illegal (i.e., not
corresponding to an encoding of the time-step) states occur in the
clock register, that the input to the circuit is correct, and that the
computation ultimately accepts.

Second, the encoding of a computation in the ground state of a
nearest-neighbor 1D Hamiltonian is based on the construction of
\cite{AGIK09}. The $n$ computational qubits are encoded in subspaces
of $n$ of the qudits (of which there are polynomially many) on the
line. The line is divided into blocks, and in each block a set of
nearest-neighbor gates is performed on the encoded qubits before the
qubits are transferred to the next block where the next set of gates
can be performed. The gate applications and the qubit transfers occur
via two-local (nearest-neighbor) operations. The construction
in~\cite{AGIK09} uses a 2-dimensional ``gate'' subspace of the qudits
to mark the position along the line where a gate is being
performed. The qudits storing the qubits on which gates have already
been performed are indicated by a two-dimensional space $\lqubit$, and
the ones on which a gate is yet to be performed are labeled
$\rqubit$. There are also one-dimensional states $\turn$ and $\lmove$:
the former marks the transition between the gate-performing steps and
the qubit-transferring steps of the computation, and the latter shifts
the qubits to the right. A two-dimensional state $\rmove$ serves to
move the active spot back to the right after $\lmove$ has moved the
qubits over one site.

Third, our main contribution is reducing the dimensionality of the qudits to 8. 
This ``leaner'' qudit construction comes at a price -- allowing the
forward/backward transitions in our Hamiltonian to be non-unique,
possibly resulting in ``illegal'' configurations of the qudit chain.
However, we can work around this problem and suppress those by adding penalty terms.
Raising the energy of states away from the allowed subspace 
allows us to use the projection lemma from \cite{KKR06},
showing that even despite the illegal transitions, 
the ground state must have a substantial overlap with the legal subspace. 
Restricted to that subspace, the illegal transitions in the Hamiltonian do not contribute to the expectation
value of the energy for a correct history state.  Therefore, the
history state of a computation that accepts with high probability can
be close to the ground state of the entire Hamiltonian, and results in a low ground-state energy. 
Of course, we also need to show a lower bound on the ground state energy for Hamiltonians 
corresponding to quantum circuits without easily-accepted witnesses.

In more detail, our dimension reduction comes from getting rid of the distinction between the two two-dimensional qudit states -- the
$\lqubit$ (done: qubits in a block that have already participated in
gate applications) and $\rqubit$ (ready: qubits that are yet to have a
gate applied to them) qubit types -- using instead just one type of
qubit $\qubit$ combined with a 1-dimensional state $\insi$, using
parity of the qubit position to distinguish between
``done''/``ready''.  We use the mapping $\lqubit \rightarrow \parity
\insi \qubit\parity$ and $\rqubit \rightarrow \parity \qubit
\insi\parity$, doubling the number of particles on the line.
Furthermore, we get rid of the $\rmove$ qubit type -- we instead use
the boundary between ``done'' and ``ready'' sequences of qubits as the
active spot.  We also will not need the $\turn$ state (used in~\cite{AGIK09})
anymore. 


\section{Background}
\label{sec:bg}

Let us begin by a general definition of the Local Hamiltonian problem:

\begin{definition}[The $d$-state $k$-Local Hamiltonian Problem]
We are given a Hamiltonian $H = H_1 + H_2 + \ldots + H_s$ on $n$ $d$-state qudits, with the matrix elements of each $H_i$ specified by $poly(n)$ bits. $H$ is $k$-local: each $H_i$ acts nontrivially on only $k$ of the $n$ qudits. We are also given two constants $a,b \in \mathbb{R}$ such that $b - a \geq 1/poly(n)$, with the promise that the smallest eigenvalue of $H$, $\lambda(H)$, is either at most $a$ or greater than $b$. We must decide if $\lambda(H) \leq a$ or $\lambda(H) > b$.
\end{definition}


As mentioned in the Introduction and shown by Kitaev~\cite{KSV02}, this problem lies in (and is complete for) the class QMA. 
\begin{definition}[QMA]
A language L is in the class QMA iff for each instance x there exists
a uniform polynomial-size quantum circuit $V_{x}$ such that 
\begin{itemize}
\item if $x \in L$, $\exists \vert \xi \rangle$, a polynomial-size
  quantum state (a witness)\\ such that $\Pr (accept(V_{x},\ket{\xi})) \geq 2/3,$ 
\item if $x \notin L$, $\forall \vert \xi \rangle$ $\Pr (accept(V_{x},\ket{\xi})) \leq 1/3$. 
\end{itemize}
\end{definition}

Previous proofs for QMA-completeness rely on a special state encoding a computation (a history state) for showing QMA-hardness of Local Hamiltonian.  A circuit is transformed into an appropriate Hamiltonian
such that a history state is a zero-eigenvector when there is a
witness to make the circuit accept.
\begin{definition}[History state]
  Let $V = U_K \cdots U_2 U_1$ be a circuit of $K$ gates on $n$ qubits. Consider a Hilbert space with $K+1$ orthogonal subspaces $\{S_t \}_{t = 0}^{K}$, each with basis $\{\ket{j_t}\}_{j=0}^{2^n-1}$ of dimension $2^n$. We define the \emph{history state} corresponding to the action of $V$ on an initial $n$-qubit state $\ket{\varphi}$ as a superposition over states coming from orthogonal spaces:
  \begin{equation}
    \ket{\eta^\varphi} = \frac{1}{\sqrt{K+1}} \sum_{t = 0}^{K} \ket{\gamma_t^{\varphi}} \label{history}
  \end{equation}
where $\ket{\gamma_t^{\varphi}} = \sum_{j=0}^{2^n-1} \ket{j_t}\bra{j} U_{t} \cdots U_2 U_1 \ket{\varphi}$ is a vector in the subspace $S_t$.
\end{definition}

Note that the Hilbert space as a whole can be bigger than the union of $S_t$'s, 
and we can write it as an orthogonal direct sum of subspaces $\left(\bigoplus_{i=0}^K S_t\right) \oplus
\cH_{rest}$, with the rest of the Hilbert space denoted $\cH_{rest}$. 

A propagation Hamiltonian can be defined to ensure that a low-energy candidate state has
the form \eqref{history}, when the state evolution satisfies a certain orthogonality condition.  
Note that for any initial $n$-qubit state $\ket{\varphi}$ and any $t\in \{0,\dots,K\}$, we have $\ket{\gamma_t^\varphi} \in S_t$.
The {\em propagation Hamiltonian} associated with the circuit $V$ is $H_{\mathrm{prop}}:= \sum_{t=0}^{K-1} H_t$ where
\begin{align}
	H_t := \sum_{j=0}^{2^n-1} \left( 
	\ket{j_{t}}\bra{j_{t}} + \ket{j_{t+1}}\bra{j_{t+1}} - U'_{t+1}  \ket{j_{t+1}}\bra{j_{t}} -  \ket{j_{t}}\bra{j_{t+1}}{U'_{t+1}}^{\dagger}
	\right),
	\label{projectorHprop}
\end{align}
with $U'_{t+1} = \sum_{k,j=0}^{2^n-1}  \ket{k_{t+1}}\bra{k}U_{t+1}\ket{j} \bra{j_{t+1}}$ acting as the unitary $U_{t+1}$ on the subspace $S_{t+1}$.

Observe that the action of $H_t$ on $\ket{\gamma_{t}^\varphi}$ and $\kets{\gamma_{t+1}^\varphi}$ is (not summing over $t$)
\begin{align}
	H_t \ket{\gamma_t^\varphi} &= \ket{\gamma_t^\varphi} - \ket{\gamma_{t+1}^\varphi}, \\
	H_t \kets{\gamma_{t+1}^\varphi} &= \kets{\gamma_{t+1}^\varphi} - \ket{\gamma_{t}^\varphi}, \nonumber 
\end{align}
since $\ket{\gamma_{t+1}^\varphi} = U'_{t+1} \sum_{j=0}^{2^n-1} \ket{j_{t+1}}\braket{j_{t}}{\gamma_{t}^\varphi}$.
It is then straightforward to verify that $\kets{ \gamma_{t}^\varphi} + \ket{
\gamma_{t+1} ^\varphi}$ is a zero-energy eigenvector of $H_t$. A history state is then also a zero eigenvector of each
$H_t$, and so a zero eigenvector of $H_{\mathrm{prop}}$.  
The propagation Hamiltonian thus serves to ``check'' the progress of the computation, 
by giving an energy penalty to all non-history states. For a specific construction 
of a QMA-complete $k$-Local Hamiltonian problem, 
it will have to be shown that $H_{\mathrm{prop}}$ can be built from operators 
that obey the chosen locality restrictions.

Let $V_x$ be a verifying circuit for an instance $x \stackrel{?}{\in} L\in QMA$,
taking as input $n-m$ ancilla qubits in the state $\ket{0}$ and an $m$-qubit state $\ket{\xi}$, 
and it has squared amplitude 2/3 on some designated output qubit if
$x\in L$, and less than 1/3 otherwise.  Kitaev's proof used a history
state of the following form.  The unitaries $U_i$ are the ones from
the original verifier circuit $V_x = U_K \dots U_1$ and are 2-local.
An extra unary clock register is used to build the structure of orthogonal subspaces $S_t$,
requiring a 2-local clock checking Hamiltonian $H_{\mathrm{clock}}$ in the Hamiltonian for distinguishing 
the subspaces $S_t$ from $\cH_{rest}$ spanned by states with illegal clock configurations. The history state
for verifying a valid witness $|\xi\rangle$ using $V_x$ is
$$|\eta\rangle = \frac{1}{\sqrt{K+1}} \sum_{t=0}^K  \left(U_t\dots U_1
\left(\kets{0}^{n-m}\otimes\ket{\xi}\right)\right)\otimes \ket{t}_{\mathrm{clock}}.$$

This history-state structure for low-energy state candidates 
is enforced by $H_{\mathrm{prop}}$ imposing energy penalties for deviating from the indicated form.
With the unary clock construction, the required locality of the terms in $H_{\mathrm{prop}}$ is 5.   
In addition, Kitaev adds two more Hamiltonian terms: $H_{\mathrm{in}}$ penalizing states
with improperly initialized ancillae
(not of the form $|\gamma_0\rangle = \kets{0}^{n-m} \otimes \ket{\xi}$), 
and $H_{\mathrm{out}}$ verifying whether the computation accepts.  
This turns out to be enough to ensure that $x\not\in L$ instances of the 5-Local Hamiltonian 
have no low-energy eigenvector.


\section{Encoding a computation in a sequence of orthogonal states of a~line of 8-dimensional qudits.} 
\label{sec:circuit}

Our goal is to encode a quantum verifier circuit $V_x$
into a 2-Local Hamiltonian instance with nearest-neighbor interactions on a
line of qudits,
satisfying certain properties.  In this section we do the first step,
transforming $V_x$
into a modified circuit $\tilde{V}_x$ that does the same computation
as $V_x$, but instead of on $n$ qubits, it acts on a line of $poly(n)$ qudits of dimension $d=8$.
All gates in $\tilde{V}_x$ are nearest-neighbor on this line, and the states occurring during
the computation are pairwise orthogonal.  This is the condition 
given in Section~\ref{sec:bg}.  
Finding the circuit $\tilde{V}_x$ with these properties allows us to define a Hamiltonian 
such that, in the case that there exists a witness on which the circuit $\tilde{V}_x$ accepts with
high probability, the history state of the computation on the witness
is a low-energy state of the Hamiltonian. Otherwise, we will be able to lower bound the ground state energy 
of this Hamiltonian.

Assume $V_x$ works on a space of $n$ qubits.  Choose a way to arrange
the qubits on a line.  The original circuit can be transformed to a circuit
$V'_x$ consisting of $R$ rounds of gates, where each round is
composed of $n-1$ nearest-neighbor gates: the first gate in a round
acts on qubits 1 and 2, the second on qubits 2 and 3, and so on. Any
quantum circuit can be recast in this fashion, by inserting swap gates
and identity gates, with a polynomial blowup increase in the number of gates.

We now convert the circuit $V'_x$ to a circuit $\tilde{V}_x$ acting on a
line of 8-state qudits arranged in $R$ blocks of $2n$ particles
each. The qudits are 8-dimensional, and we will utilize some of the 8 states 
as data-carriers (holding qubits the computation acts on). The rest   
will guarantee the orthogonality conditions and the proper progress of the computation. 
At any time during the computation, we want exactly $n$ of the qudits to be 
in the ``data-holding'' states, and we simply call them qubits. 
Initially, all of the $n$ qubits are located in the first block of particles. 
After each round of gates from $V'_x$ is carried out, the
qubits are transferred to the next block of $2n$ particles where the
next set of gates from $V'_x$ can be performed.  


\vspace*{12pt}
\noindent
\begin{claim}
  Given a QMA verifier circuit $V_x$ on $n$ qubits,  an equivalent QMA
  verifier circuit $\tilde{V}_x$ can be efficiently computed such that
  $\tilde{V}_x$ operates
  on $2nR$ 8-state qudits on a line, only uses nearest
  neighbor gates, and such that the states occurring during the
  computation are pairwise orthogonal.  
\end{claim}

\vspace*{12pt}
\noindent

In the rest of this Section we describe the sequence of orthogonal states that appear in the computation on the qudit line.
Later, in Section \ref{sec:hamiltonian} we present the positive semidefinite 2-local Hamiltonian whose ground state
is the uniform superposition over states from this desired sequence.

Let us choose the Hilbert space of each particle as an
orthogonal direct sum: $\cH_8=\insi \oplus \lmove
\oplus \blnk \oplus \dead \oplus \qubit \oplus \gate$.
The subspaces denoted $\insi, \lmove, \blnk$, and $\dead$ are 1-dimensional. 
Then we have 2-dimensional subspaces $\qubit$ and $\gate$ , each designed to hold a state of a qubit (specified
by two complex numbers $a_{0}$ and $a_{1}$, with $a_{0}$, $a_{1}$ $\in
\mathbb{C}$ and $\vert a_{0} \vert^2 + \vert a_{1} \vert^2 = 1$).
We label the basis vectors of these 2-dimensional subspaces $\kets{\qubit^{(s)}}$ and $\kets{\gate^{(s)}}$ with $s=0,1$. 
A qudit in the state $\sum_{s=0}^{1} a_{s} \kets{\qubit^{(s)}}$
or $\sum_{s=0}^{1} a_{s} \kets{\gate^{(s)}}$ is then said to have the \emph{qubit content} $a_{0} \ket{0}+a_1 \ket{1}$. 

When we label a qudit by one of the symbols $\{\insi, \lmove, \blnk,
\dead, \qubit, \gate\}$, we mean that its state belongs to a particular subspace of $\cH_8$. Such labeling of the whole chain
defines a \emph{configuration}. The Hilbert space of the qudit chain thus decomposes into orthogonal subspaces 
indexed by configurations. We can choose a basis for the Hilbert
space of the entire system as a tensor product of $2nR$ (one for each
site) of the basis vectors $\left\{ \kets{\insi}, \kets{\lmove}, \kets{\blnk}, \kets{\dead}, \kets{\qubit^{(0)}}, \kets{\qubit^{(1)}}, \kets{\gate^{(0)}}, \kets{\gate^{(1)}}
\right\}$. The \emph{state} of
the system is a vector in the span of the basis vectors. 

Let us now construct a sequence of configurations, corresponding to the progression of a computation with the circuit $\tilde{V}_x$.
We view the qudit chain as $R$ blocks of length $2n$ and mark their boundaries $\bdry$. To highlight the parity of the sites, we also draw $\parity$ after every even, non-boundary site.

In the initial configuration, the first block holds qubits at odd-numbered sites, interspersed with $\insi$s. The rest of the chain consists of $\blnk$s:
\begin{align}
\bdry \underbrace{\gate \insi \parity \qubit \insi \parity \mgdots \parity \qubit \insi \parity \qubit \blnk }_\text{the first block of length $2n$}
\bdry \blnk \blnk \parity \blnk \blnk \parity \mgdots \label{initialconfiguration}
\end{align}
The qubit content of the $\gate$ and $\qubit$ sites carries the initial $n$-qubit input to the
circuit $V'_x$ (the ancillae and the witness). Each step of the computation is a 2-local unitary operation
applied to two adjacent particles, resulting in a change of configuration (building up an orthogonal sequence),
and possibly a change in the state of the qubit content (doing the computation). 
Let us now write the rules for building up the circuit $\tilde{V}_x$. 

\begin{table} 
\begin{enumerate}
\item \label{rule:gate} $\gate \parity \qubit \longleftrightarrow
  U_m(\qubit \parity \gate) $
	  performs a two-qubit gate $U_m$ (location-dependent) on the qubit content of the two particles, 
	  while shifting the active site to the right.
\item \label{rule:move} 
\begin{enumerate}
\item \label{rule:move1} $ \parity \gate \insi \parity  \longleftrightarrow \parity  \insi \gate \parity $ 
	  moves an ``active'' qubit $\gate$ to the right (not near a block boundary),
\item \label{rule:move2} $ \bdry \gate \insi \parity  \longleftrightarrow \bdry \dead \gate \parity  $ 
	  	is applicable when a block boundary is to the left of it,
\item  \label{rule:move3} $\parity  \gate \blnk \bdry \longleftrightarrow  \parity  \insi \gate \bdry $ 
	  	is applicable when a block boundary is in front of it.
\end{enumerate}
\item \label{rule:moveq} 
\begin{enumerate}
  
	\item \label{rule:moveq2} $ \dead \parity \qubit \insi \bothp\qubit   \longleftrightarrow \dead \parity \dead \qubit \bothp \qubit  $ 
	  moves the leftmost qubit (not after a boundary),  
	\item \label{rule:moveq1} $ \qubit \bothp \qubit \insi \bothp\qubit   
		\longleftrightarrow \qubit \bothp \insi \qubit \bothp\qubit  $ 
	  moves a qubit $\qubit$ to the right, only noting correct parity, 
	  regardless of the boundary location. We denote this using the symbol $\bothp$.
	    
	\item \label{rule:moveq3} $ \qubit \bothp \qubit \blnk \parity \blnk   \longleftrightarrow \qubit \bothp \insi \qubit \parity \blnk  $ 
	  moves the rightmost qubit (not before a boundary). 
  \item \label{rule:moveq4}	$ \dead \bothp \qubit \blnk \bothp \blnk   \longleftrightarrow \dead \bothp \dead \qubit \bothp \blnk  $ 
	  a special rule ensuring that if there is a {\em single} qubit in the chain, it can still
          move. Rule 3(d) does not actually apply to any legal configuration.
  \end{enumerate}
    
\item \label{rule:make} 
\begin{enumerate}
\item \label{rule:make1} $\gate \bdry \blnk \blnk \longleftrightarrow \qubit \bdry \lmove \blnk$ creates a left-moving pusher $\lmove$ 
			at the front near a boundary $\bdry$.
\item \label{rule:make2} $\qubit \parity \blnk \blnk \longleftrightarrow \qubit \parity \lmove \blnk$ 
			introduces $\lmove$ when away
                        from a block boundary.
\end{enumerate}
     
\item \label{rule:left} 
\begin{enumerate}
\item \label{rule:left1} $\ppp \qubit \bothp \lmove \longleftrightarrow \ppp\lmove \bothp \qubit$ 
			pushes $\lmove$ left and a qubit to the right (not caring for the boundary).
\item \label{rule:left2} $\bothp \insi  \lmove \bothp \longleftrightarrow \bothp \lmove \insi\bothp$ 
			does the same with $\lmove$ and $\insi$, at
                        locations with this parity.
\end{enumerate}

\item \label{rule:kill} 
\begin{enumerate}
\item \label{rule:kill1}$\dead \lmove \bdry \qubit \longleftrightarrow \dead \dead \bdry \gate$ kills the pusher $\lmove$ 
			at the left end of the qubits at a boundary $\bdry$, 
			changing the last qubit to $\gate$, allowing the next round of gate applications to begin. 
\item \label{rule:kill2}$\dead \lmove \parity \qubit \longleftrightarrow \dead
  \dead \parity \qubit$ simply kills the pusher, when away from the boundary.
  $\lmove$.
\end{enumerate}
\end{enumerate}
\vspace{5pt}
\caption{The transition rules, which together with a carefully chosen initial state \eqref{initialconfiguration} define the 2-local gates of the circuit $\tilde{V}_x$. 
Note that some of these rules are 2-local, some 3-local and some even 4-local, which helps them identify their 
intended locations uniquely. However, the transformations themselves are only 2-local.
See also Table~\ref{example} for an example of a progression of configurations and the unique applicability of these rules. 
We will later write a Hamiltonian $H_{\textrm{prop}}$ with only 2-local terms checking these transitions.}
\label{table:rules}
\end{table}

We choose a list of transition rules (for configurations) and list them in Table~\ref{table:rules}. Each rule connects configurations that differ in two particular neighboring spots, and are connected by a 2-local unitary transformation. The sequence of these transformations (as applied sequentially to the initial configuration) defines the 2-local gates of the circuit $\tilde{V}_x$. This assignment of unitaries is unique by construction, as we choose the transition rules so that for any configuration arising from the initial one, there is always exactly one rule that possibly applies to it (see also Table~\ref{example} for a part of the sequence of configurations for $n=3$).
We ensure this uniqueness by rules involving up to 4 particles in the rules. However, in Section~\ref{sec:hamiltonian} 
we will write a Hamiltonian made from 2-local terms that checks\footnote{To ``check'' a transition means adding an energy
  penalty to terms that do not have the same amplitude for both of the
  states involved in the transition.} these transitions.



Let us explain the logic behind the rules. Rule~\ref{rule:gate} applies the unitary from the modified, nearest-neighbor circuit $V'_x$.  The rest of the rules ensure the orthogonalization and locality properties.
Initially, the qubits are placed at the odd sites, separated by $\insi$s. 
If we want them to interact, we have to move them together, 
which is what rules \ref{rule:move} and \ref{rule:moveq} do. The nearest-neighbor gates from 
$V'_x$ are then performed at the $\gate \qubit$ junctions using rule~\ref{rule:gate}. The $\blnk $ label marks sites 
that the computation hasn't reached yet, while the $\dead$ sites will
not be used again.
The $\lmove$ (a pusher state) serves to move the qubits to the right. 

The computation can be divided into $R$ ``rounds'', each corresponding to the application of a ``round'' of gates from $V'_x$, and then moving the qubit block $2n$ positions to the right. 
Let us look at the two phases of a round of computation in detail, referring to Table~\ref{example} (a $n=3$ qubit example).  
 
The goal of the first phase of the computation is gate application.
It involves rules \ref{rule:move} and \ref{rule:gate}, moving the $\gate$ qubit from the left end of the chain
while applying the gates from a given ``round''.
When $\gate$ reaches the front end of the chain, rule~\ref{rule:make} creates a the ``pusher'' state $\lmove$. 
After $2n-1$ applications of rule~\ref{rule:left}, the pusher gets to the left end of the qubit sequence, where it disappears through rule~\ref{rule:kill}.
This first phase thus moves $\gate$ $n$ times, makes $n-1$ gate applications, adds $1$ pusher creation, $2n-1$ pushes and $1$ killing of $\lmove$, altogether making $4n$ steps.

The second phase (which is repeated $n-1$ times) moves the qubits to
the right until they are all within the next block. It takes $n$ applications
of rule~\ref{rule:moveq} to move all the qubits one step to the right.
Then we create the pusher $\lmove$, move it to the left ($2n-1$ steps)
and kill it. Altogether, this takes $3n+1$ steps. If we now are not at
the boundary, the second phase repeats. If we are at a block 
boundary~$\bdry$, the second phase concludes, and the ``round'' of computation
concludes as well, as all the qubits have now moved $2n$ positions to
the right. A new ``round'' of computation (with the particle $\gate$
starting to move) starts according to rule~\ref{rule:move}.  Summing
it up, a whole ``round'' of computation consists of $4n + (n-1)(3n+1) =
3n^2+2n-1$ steps.  During each ``round'', $n-1$ gates from $V'_x$ are
applied and the qubits are moved over to the next block of
qudits. This happens for each of the first $R-1$ blocks. In the last
block, after the gates are applied, the computation comes to a halt in
the state: $\bdry \dead^{2n(R-1)} \bdry \dead \qubit \parity \insi
\qubit \parity \mgdots \parity \insi \qubit \parity \insi \gate
\bdry$.
Also, without loss of generality, we take all the gates in the very first round to be identities. This allows us to verify that the ancilla qubits (laid out on the left of the qubit sequence) all start out in the correct state $\ket{0}$. 

The entire computation with $(R-1)$ regular rounds and a last round with $2n$ steps (until $\gate$ reaches the right end) together take 
$K= (R-1)(3n^2+2n-1)+2n$ steps, corresponding to $K+1$ configurations of the qudits. Also note that a configuration is never
repeated in the course of the computation -- all of the $K+1$
configurations are distinct, and therefore orthogonal. 

\begin{table}
\begin{center}

\begin{tabular}{ll}
$\mgdots \dead \dead \bdry \gate \insi \parity \qubit \insi \parity \qubit \blnk \bdry \blnk \blnk \parity \blnk \blnk \parity \blnk \blnk \bdry \blnk \blnk \mgdots$&\quad rule \ref{rule:move}\\
$\mgdots \dead \dead \bdry \dead \gate \parity \qubit \insi \parity \qubit \blnk \bdry \blnk \blnk \parity \blnk \blnk \parity \blnk \blnk \bdry \blnk \blnk \mgdots$&\quad rule \ref{rule:gate} \\
$\mgdots \dead \dead \bdry \dead \qubit\parity \gate  \insi \parity \qubit \blnk \bdry \blnk \blnk \parity \blnk \blnk \parity \blnk \blnk \bdry \blnk \blnk \mgdots$&\quad rule \ref{rule:move}\\ 
$\mgdots \dead \dead \bdry \dead \qubit\parity \insi  \gate \parity \qubit \blnk \bdry \blnk \blnk \parity \blnk \blnk \parity \blnk \blnk \bdry \blnk \blnk \mgdots$&\quad rule \ref{rule:gate} \\
$\mgdots \dead \dead \bdry \dead \qubit\parity \insi  \qubit\parity \gate  \blnk \bdry \blnk \blnk \parity \blnk \blnk \parity \blnk \blnk \bdry \blnk \blnk \mgdots$&\quad rule \ref{rule:move} \\
$\mgdots \dead \dead \bdry \dead \qubit\parity \insi  \qubit\parity \insi  \gate \bdry \blnk \blnk \parity \blnk \blnk \parity \blnk \blnk \bdry \blnk \blnk \mgdots$&\quad rule \ref{rule:make} \\
$\mgdots \dead \dead \bdry \dead \qubit\parity \insi  \qubit\parity \insi  \qubit\bdry \lmove\blnk \parity \blnk \blnk \parity \blnk \blnk \bdry \blnk \blnk \mgdots$&\quad rule \ref{rule:left} \\
$\mgdots \dead \dead \bdry \dead \qubit\parity \insi  \qubit\parity \insi  \lmove\bdry \qubit\blnk \parity \blnk \blnk \parity \blnk \blnk \bdry \blnk \blnk \mgdots$&\quad rule \ref{rule:left} \\
$\mgdots \dead \dead \bdry \dead \qubit\parity \insi  \qubit\parity \lmove \insi \bdry \qubit\blnk \parity \blnk \blnk \parity \blnk \blnk \bdry \blnk \blnk \mgdots$&\quad rule \ref{rule:left} \\
$\mgdots \dead \dead \bdry \dead \qubit\parity \insi  \lmove\parity \qubit \insi \bdry \qubit\blnk \parity \blnk \blnk \parity \blnk \blnk \bdry \blnk \blnk \mgdots$&\quad rule \ref{rule:left} \\
$\mgdots \dead \dead \bdry \dead \qubit\parity \lmove \insi \parity \qubit \insi \bdry \qubit\blnk \parity \blnk \blnk \parity \blnk \blnk \bdry \blnk \blnk \mgdots$&\quad rule \ref{rule:left} \\
$\mgdots \dead \dead \bdry \dead \lmove\parity \qubit \insi \parity \qubit \insi \bdry \qubit\blnk \parity \blnk \blnk \parity \blnk \blnk \bdry \blnk \blnk \mgdots$&\quad rule \ref{rule:kill} \\
$\mgdots \dead \dead \bdry \dead \dead \parity \qubit \insi \parity \qubit \insi \bdry \qubit\blnk \parity \blnk \blnk \parity \blnk \blnk \bdry \blnk \blnk \mgdots$&\quad rule \ref{rule:moveq} \\
$\mgdots \dead \dead \bdry \dead \dead \parity \dead  \qubit\parity \qubit \insi \bdry \qubit\blnk \parity \blnk \blnk \parity \blnk \blnk \bdry \blnk \blnk \mgdots$&\quad rule \ref{rule:moveq} \\
$\mgdots \dead \dead \bdry \dead \dead \parity \dead  \qubit\parity \insi  \qubit\bdry \qubit\blnk \parity \blnk \blnk \parity \blnk \blnk \bdry \blnk \blnk \mgdots$&\quad rule \ref{rule:moveq} \\
$\mgdots \dead \dead \bdry \dead \dead \parity \dead  \qubit\parity \insi  \qubit\bdry \insi \qubit\parity \blnk \blnk \parity \blnk \blnk \bdry \blnk \blnk \mgdots$&\quad rule \ref{rule:make} \\
$\mgdots \dead \dead \bdry \dead \dead \parity \dead  \qubit\parity \insi  \qubit\bdry \insi \qubit\parity \lmove\blnk \parity \blnk \blnk \bdry \blnk \blnk \mgdots$&\quad rule \ref{rule:left} \\
$\mgdots \dead \dead \bdry \dead \dead \parity \dead  \qubit\parity \insi  \qubit\bdry \insi \lmove\parity \qubit\blnk \parity \blnk \blnk \bdry \blnk \blnk \mgdots$&\quad rule \ref{rule:left} \\
$\mgdots \dead \dead \bdry \dead \dead \parity \dead  \qubit\parity \insi  \qubit\bdry \lmove\insi \parity \qubit\blnk \parity \blnk \blnk \bdry \blnk \blnk \mgdots$ &\quad rule \ref{rule:left}\\
$\mgdots \dead \dead \bdry \dead \dead \parity \dead  \qubit\parity \insi  \lmove\bdry \qubit\insi \parity \qubit\blnk \parity \blnk \blnk \bdry \blnk \blnk \mgdots$&\quad rule \ref{rule:left} \\
$\mgdots \dead \dead \bdry \dead \dead \parity \dead  \qubit\parity \lmove \insi \bdry \qubit\insi \parity \qubit\blnk \parity \blnk \blnk \bdry \blnk \blnk \mgdots$&\quad rule \ref{rule:left} \\
$\mgdots \dead \dead \bdry \dead \dead \parity \dead  \lmove\parity \qubit \insi \bdry \qubit\insi \parity \qubit\blnk \parity \blnk \blnk \bdry \blnk \blnk \mgdots$&\quad rule \ref{rule:kill} \\
$\mgdots \dead \dead \bdry \dead \dead \parity \dead  \dead \parity \qubit \insi \bdry \qubit\insi \parity \qubit\blnk \parity \blnk \blnk \bdry \blnk \blnk \mgdots$&\quad rule \ref{rule:moveq} \\
$\mgdots \dead \dead \bdry \dead \dead \parity \dead  \dead \parity \dead  \qubit\bdry \qubit\insi \parity \qubit\blnk \parity \blnk \blnk \bdry \blnk \blnk \mgdots$ &\quad rule \ref{rule:moveq}\\
$\mgdots \dead \dead \bdry \dead \dead \parity \dead  \dead \parity \dead  \qubit\bdry \insi \qubit\parity \qubit\blnk \parity \blnk \blnk \bdry \blnk \blnk \mgdots$&\quad rule \ref{rule:moveq} \\
$\mgdots \dead \dead \bdry \dead \dead \parity \dead  \dead \parity \dead  \qubit\bdry \insi \qubit\parity \insi \qubit\parity \blnk \blnk \bdry \blnk \blnk \mgdots$ &\quad rule \ref{rule:make}\\
$\mgdots \dead \dead \bdry \dead \dead \parity \dead  \dead \parity \dead  \qubit\bdry \insi \qubit\parity \insi \qubit\parity \lmove\blnk \bdry \blnk \blnk \mgdots$ &\quad rule \ref{rule:left}\\
$\mgdots \dead \dead \bdry \dead \dead \parity \dead  \dead \parity \dead  \qubit\bdry \insi \qubit\parity \insi \lmove\parity \qubit\blnk \bdry \blnk \blnk \mgdots$ &\quad rule \ref{rule:left}\\
$\mgdots \dead \dead \bdry \dead \dead \parity \dead  \dead \parity \dead  \qubit\bdry \insi \qubit\parity \lmove\insi \parity \qubit\blnk \bdry \blnk \blnk \mgdots$ &\quad rule \ref{rule:left}\\
$\mgdots \dead \dead \bdry \dead \dead \parity \dead  \dead \parity \dead  \qubit\bdry \insi \lmove\parity \qubit\insi \parity \qubit\blnk \bdry \blnk \blnk \mgdots$&\quad rule \ref{rule:left} \\
$\mgdots \dead \dead \bdry \dead \dead \parity \dead  \dead \parity \dead  \qubit\bdry \lmove\insi \parity \qubit\insi \parity \qubit\blnk \bdry \blnk \blnk \mgdots$ &\quad rule \ref{rule:left}\\
$\mgdots \dead \dead \bdry \dead \dead \parity \dead  \dead \parity \dead  \lmove\bdry \qubit\insi \parity \qubit\insi \parity \qubit\blnk \bdry \blnk \blnk \mgdots$ &\quad rule \ref{rule:move}\\
$\mgdots \dead \dead \bdry \dead \dead \parity \dead  \dead \parity \dead  \dead \bdry \gate \insi \parity \qubit\insi \parity \qubit\blnk \bdry \blnk \blnk \mgdots$& 
\end{tabular}

\end{center}
\vspace{5pt}
\caption{The configurations occurring in one cycle of the computation with $n=3$ qubits. The rules whose application brings the state to the next one are listed on the right.}
\label{example}
\end{table}


\subsection{Legal configurations}
\label{sec:legal}

\newcommand{\lxx}{x} 
\newcommand{\lxd}{D$_x$} 
\newcommand{\lDD}{D} 
\newcommand{\lAleft}{A$_x$} 
\newcommand{\lGleft}{A$_1$}  
\newcommand{\lAp}{A$_{p}$} 
\newcommand{\lGright}{A$_2$} 
\newcommand{\lAright}{A$_u$} 
\newcommand{\lRR}{R} 
\newcommand{\lru}{R$_u$} 
\newcommand{\luu}{u} 

At the moment, we are interested only in the (legal) configurations that we want to appear during a computation. Of course, the whole Hilbert space is much larger, containing many other states. We will call those illegal, and want them to be ``detectable''. For now, we will not deal with these other states until Section \ref{sec:penalty}.

Let the set of {\em legal} configurations $C_0,\dots,C_K$ be the $K+1$
configurations that can be obtained by applying the rules in Table~\ref{table:rules} starting with the
initial configuration~\eqref{initialconfiguration}. The legal configurations correspond to the $K+1$ (including the initial state)
intermediate computational states generated by the circuit $\tilde{V}_x$. We call all other
configurations {\em illegal}.

We will now look at the properties shared by the legal configurations. 
It will be convenient to look at pairs of particles at locations $(2i-1,2i)$ and $(2i+1,2i+2)$.
Table~\ref{table:activity} lists the allowed pairs of symbols and
which ones can be adjacent to each other.  The pairs play the roles
of ``dead'' (labeled x, particles not to be used anymore), ``done'' (labeled D, qubits to
the left of the active site), ``active'' (labeled A, the active site), ``ready''
(labeled R, qubits to the right of the active site) and ``unborn'' (labeled u, ``unborn'' particles, not
used yet) from the construction in \cite{AGIK09}. There, the legal states were of the form $\textrm{(\lxx\,$\cdots$\,\lxx) (\lDD\,$\cdots$\,\lDD) A (\lRR\,$\cdots$\,\lRR) (\luu\,$\cdots$\,\luu)}$, with a single active site. Here $\textrm{(z\,$\cdots$\,z)}$ stands for a variable-length string made from the letter ``z''.

Connecting subsequent pairs according to the rules listed in Table~\ref{table:activity} imposes a particular form for the legal states (brackets indicate variable-length, possibly empty substrings)
\begin{align}
	 \textrm{(\lxx\,$\cdots$\lxx)$[qubits]$(\luu\,$\cdots$\luu)} \label{properform1}
\end{align}
where [{\em qubits}] is a {\em nonzero} string of the form
  \begin{align}
	&\hskip21mm
		\textrm{\lAleft}
		\textrm{(\lRR\,$\cdots$\lRR)\, \lru}    \label{legalqubitsstart}\\ 
	&\, \, \, \textrm{(\lxd \lDD\,$\cdots$\lDD)}
		\textrm{\lGleft}
		\textrm{(\lRR\,$\cdots$\lRR)\, \lru} \label{legalqubits2}\\
	&\textrm{\lxd\, (\lDD\,$\cdots$\lDD)}	
		\textrm{\lAp}
		\textrm{(\lRR\,$\cdots$\lRR)\, \lru}\\
	&\textrm{\lxd\, (\lDD\,$\cdots$\lDD)}
		\textrm{\lGright}
		\textrm{(\lRR\,$\cdots$\lRR \lru)}\\
	& \textrm{\lxd\, (\lDD\,$\cdots$\lDD)}
		\textrm{\lAright}
		\hskip20.5mm\\
	&\hskip26mm
		\textrm{(\lRR $\cdots$\lRR)\, \lru}\\
	&\textrm{\lxd\, (\lDD\,$\cdots$\lDD)}
		\hskip4.5mm
		\textrm{(\lRR\,$\cdots$\lRR)\, \lru}\\
	&\textrm{\lxd\, (\lDD\,$\cdots$\lDD)}  .
	\hskip23mm \label{legalqubitsend}
   \end{align}
The first five options involve an ``active'' pair, while the last three have no ``active'' pair in them. Furthermore, note that the whole [{\em qubits}] string cannot be empty, because the rightmost particle of the whole chain cannot be $\dead$, 
the leftmost one cannot be $\blnk$ and the combination $\dead \blnk$ is illegal. Next, for legal configurations, the number of particles holding qubits needs to be exactly $n$.


Let us have a closer look at the legal [{\em qubits}] strings with an active site ($\gate$ or $\lmove$), which 
translates to a single active pair (\lAleft, \lGleft, \lAp, {\lGright} or \lAright).
One example is
$\bdry \dead \qubit \parity \gate \insi \parity \qubit \blnk \bdry$
(which is of the type \lxd \lGleft \lru).
In the case there is no {\lxd} pair, the active pair has to be 
\lAleft \eqref{legalqubitsstart} \ or \lGleft \eqref{legalqubits2} -- an example is the sequence
$\bdry \dead \gate \parity \qubit \insi \parity \qubit \blnk
\bdry$ (with pairs \lAleft  \lRR \lru).
In the case there is no {\lru} pair, the active pair has to be of the
\lGright\ or \lAright,
as in e.g. $\bdry \dead \qubit \parity \insi \qubit \parity \insi \qubit \bdry \lmove \blnk \parity$
(this is \lxd \lDD \lDD \lAright).

The other three types of legal substrings [{\em qubits}] do not have an active pair.
First, we could have a done qubit pair on the right end as in 
$\parity \dead \qubit \bdry \insi \qubit \parity \insi \qubit \parity$ 
(simply \lxd \lDD \lDD \ without any \lRR's).
Second, observe that two neighboring $\qubit$ particles can appear at positions $(2i,2i+1)$, 
when coming from two consecutive pairs as in $\parity \dead \qubit \parity \insi \qubit
\bdry \qubit \blnk \parity$ (read as \lxd \lDD \lru).
Finally, it is possible to have a ready qubit pair on the left end as in 
$\parity \qubit \insi \bdry \qubit \insi \parity \qubit \blnk \parity$ 
(simply \lRR \lRR \lru \ without any \lDD's). 

The location of the [{\em qubits}] substring matters.
For a legal configuration with a $\gate$
symbol, the string [{\em qubits}] must fit exactly between two block
boundaries as $\bdry [qubits] \bdry$ (see Table~\ref{example}). On the other hand,
the string [{\em qubits}] without the symbol $\gate$ always has runs across a block
boundary somewhere.  These two properties later help us check that we
do not have too few or too many qubits or whether the qubits are
properly aligned between the boundaries, ruling out illegal but 
locally undetectable states.

\begin{table}
\begin{center}
\hskip6.3cm\,
\begin{tabular}{|c|}
\hline
	\hskip2cm {\em allowed to be followed by} \hskip2cm \,
\end{tabular}\\
\begin{adjustbox}{width={\textwidth},totalheight={\textheight},keepaspectratio}
\begin{tabular}{|l|l|c|c|c|c|c|c|c|c|c|c|c|c|c|c|c|}
\hline
	{\em property}  & {\em symbol pair} & 
	\lxx & \lxd  & \lDD & \lAleft &
        \lGleft &  \lAp & \lGright & \lAright &
        \lRR &  \lru & \luu
\\
\hline
dead: \lxx
		& $\dead \dead$ & 
		\ok & \ok  &  & \ok & \ok  &  & & &\ok & \ok &\\
\hline
dead+done: \lxd 
		& $\dead \qubit$ & 
		 &  & \ok &  & \ok & \ok & \ok & \ok & \ok & \ok & \ok\\
\hline
done: \lDD
		& $\insi \qubit$ & 
		 &  & \ok &  & \ok  & \ok  &\ok  & \ok & \ok & \ok & \ok\\		
\hline
active leftmost: \lAleft
		& $\dead \gate$,\, $\dead\lmove$ \,  &
		 &  &  &  &  &  & & & \ok & \ok & \\
\hline
active gate 1: \lGleft
		& $\gate \insi$\,  &
		 &  &  &  &  &  & & & \ok & \ok & \\
\hline
active pusher: \lAp
		& $\insi \lmove$,\,
		$\lmove \insi$\, & 
		 &  &  &  &  &  & & & \ok & \ok &\\
\hline
active gate 2: \lGright
		& $\insi \gate$\,  &
		 &  &  &  &  &  & & & \ok & \ok  & \ok \\
\hline
active rightmost: \lAright
		& $\gate \blnk$,\, 
		 $\lmove\blnk$\, &
		 &  &  &  &  &  & & & & & \ok\\
\hline
ready: \lRR
		& $\qubit \insi$ & 
		 &  &  &  &  &  & & & \ok & \ok &\\
\hline
ready+unborn: \lru
		& $\qubit \blnk$ & 
		 &  &  &  &  &  & & & & & \ok\\
\hline
unborn: \luu
		& $\blnk \blnk$ &  
		&  &  &  &  &  & & & & & \ok\\
\hline
\end{tabular}
\end{adjustbox}
\end{center}
\vspace{5pt}
\caption{Building up the legal configuration structure from pairs of symbols (unlisted symbol pairs do not appear in legal configurations). 
We list symbol pairs allowed at positions $(2i-1,2i)$ 
and their designated followups at positions $(2i+1,2i+2)$.
Note the mirror symmetry of the table across the antidiagonal.
The allowed configurations of the whole chain must then have form
(\lxx \,$\cdots$\lxx)$[qubits]$(\luu\,$\cdots$\luu),
with a substring [{\em
  qubits}] given by \eqref{legalqubitsstart}-\eqref{legalqubitsend}, with at most one active pair.
Further restrictions come into play from considering the block boundary locations (see Table~\ref{table:legalpairs}) 
the number of ``qubits'' and their proper alignment with respect to the block boundaries.
}
\label{table:activity}
\end{table}

\begin{table} 
\begin{enumerate}
\item $\dead \bdry \gate \insi \parity (\qubit \insi)^{n-2} \parity \qubit \blnk \bdry (\blnk \blnk)^{n} \bdry \blnk$ (\ref{rule:move2})
\item $\dead \bdry \dead \gate \parity (\qubit \insi)^{n-2} \parity \qubit \blnk \bdry (\blnk \blnk)^{n} \bdry \blnk$  (\ref{rule:gate})
\item For $i$ from 0 to $n-3$:
  \begin{enumerate}
  \item $\dead \bdry \dead \qubit \parity (\insi \qubit)^{i} \parity
    \gate \insi \parity (\qubit \insi)^{n-i-3} \parity \qubit \blnk
    \bdry (\blnk \blnk)^{n} \bdry \blnk$ (\ref{rule:move1})
  \item $\dead \bdry \dead \qubit \parity (\insi \qubit)^{i} \parity
    \insi \gate \parity (\qubit \insi)^{n-i-3} \parity \qubit \blnk
    \bdry (\blnk \blnk)^{n} \bdry \blnk$ (\ref{rule:gate})
  \end{enumerate}
\item $\dead \bdry \dead \qubit \parity (\insi \qubit)^{n-2} \parity
  \gate \blnk \bdry (\blnk \blnk)^{n} \bdry \blnk$ (\ref{rule:move3})
\item \label{evolve:end} $\dead \bdry \dead \qubit \parity (\insi
  \qubit)^{n-2} \parity \insi \gate \bdry (\blnk \blnk)^{n} \bdry
  \blnk$ (\ref{rule:make1})
\item $\dead \bdry \dead \qubit \parity (\insi \qubit)^{n-1} \bdry
  \lmove \blnk \parity (\blnk \blnk)^{n-1} \bdry \blnk$ (\ref{rule:left1})
\item \label{evolve:loopa} For $j$ from 0 to $n-2$:
  \begin{enumerate}
  \item For $k$ from 0 to $n-2$:
  \begin{enumerate}
    \item $\dead \bdry (\dead \dead)^{j} \parity \dead \qubit \parity
      (\insi \qubit)^{n-k-2} \parity \insi \lmove \parity (\qubit
      \insi)^{k} \parity \qubit \blnk \parity (\blnk \blnk)^{n-j-1}
      \bdry \blnk$ (\ref{rule:left2})
    \item $\dead \bdry (\dead \dead)^{j} \parity \dead \qubit \parity
      (\insi \qubit)^{n-k-2} \parity \lmove \insi \parity (\qubit
      \insi)^{k} \parity \qubit \blnk \parity (\blnk \blnk)^{n-j-1}
      \bdry \blnk$ (\ref{rule:left1})
    \end{enumerate}
  \item $\dead \bdry (\dead \dead)^{j} \parity \dead \lmove \parity
    (\qubit \insi)^{n-1} \parity \qubit \blnk \parity (\blnk
    \blnk)^{n-j-1} \bdry \blnk$ (\ref{rule:kill2})
  \item $\dead \bdry (\dead \dead)^{j} \parity \dead \dead \parity
    (\qubit \insi)^{n-1} \parity \qubit \blnk \parity (\blnk
    \blnk)^{n-j-1} \bdry \blnk$ (\ref{rule:moveq2})
  \item For $l$ from 0 to $n-3$: \\
   $\dead \bdry (\dead \dead)^{j+1} \parity \dead \qubit \parity (\insi
  \qubit)^{l} \parity (\qubit \insi)^{n-l-2} \parity \qubit
  \blnk \parity (\blnk \blnk)^{n-j-1} \bdry \blnk$ (\ref{rule:moveq1})
\item     $\dead \bdry (\dead \dead)^{j+1} \parity \dead \qubit \parity (\insi
  \qubit)^{n-2} \parity \qubit
  \blnk \parity (\blnk \blnk)^{n-j-1} \bdry \blnk$  (\ref{rule:moveq3})

\item $\dead \bdry (\dead \dead)^{j+1} \parity \dead \qubit \parity
  (\insi \qubit)^{n-2} \parity \insi \qubit \parity (\blnk
  \blnk)^{n-j-1} \bdry \blnk$   (\ref{rule:make2}) 
  \item $\dead \bdry (\dead \dead)^{j+1} \parity \dead \qubit \parity
    (\insi \qubit)^{n-2} \parity \insi \qubit \parity \lmove
    \blnk \parity (\blnk \blnk)^{n-j-2} \bdry \blnk$ (\ref{rule:left1}) 
  \end{enumerate}

  \item \label{evolve:loopb} For $i$ from 0 to $n-2$:
\begin{enumerate}  
  \item $\dead \bdry (\dead \dead)^{n-1} \parity \dead \qubit \parity
    (\insi \qubit)^{n-i-2} \parity \insi \lmove \parity (\qubit
    \insi)^{i} \parity \qubit \blnk \bdry \blnk$ (\ref{rule:left2}) 
  \item $\dead \bdry (\dead \dead)^{n-1} \parity \dead \qubit \parity
    (\insi \qubit)^{n-i-2} \parity \lmove \insi \parity (\qubit
    \insi)^{i} \parity \qubit \blnk \bdry \blnk$ (\ref{rule:left1}) 
\end{enumerate}

\item $\dead \bdry (\dead \dead)^{n-1} \parity \dead \lmove \bdry
  (\qubit \insi)^{n-1} \parity \qubit \blnk \bdry \blnk$ (\ref{rule:kill1}) 
\item $\dead \bdry (\dead \dead)^{n} \bdry \gate \insi \parity (\qubit \insi)^{n-2} \parity \qubit \blnk \bdry \blnk$
\end{enumerate}
\vspace{5pt}
\caption{The general form of the sequence of legal configurations in one round of
  computation.  The middle block boundary is not shown in
  steps~\ref{evolve:loopa} and \ref{evolve:loopb}.  The full
  computation ends at Step~\ref{evolve:end} without the trailing circles. A particular example for $n=3$ is shown in Table~\ref{example}.} 
\label{table:evolve}
\end{table}


\section{The Hamiltonian}
\label{sec:hamiltonian}

We aim to construct a Hamiltonian corresponding
to a circuit $\tilde{V}_x$ such that the ground state energy of the
Hamiltonian is $E\leq a$ for `yes' instances ($x \in L$) and $E\geq b$
for the `no' instances, where $a$ and $b$ have a $1/poly(n)$
separation.  We will show the history state of the computation on the witnesses for `yes' instances
has a low energy.
Our Hamiltonian is a sum of four terms:
$$H := J_{\mathrm{in}}H_{\mathrm{in}} +
J_{\mathrm{prop}}H_{\mathrm{prop}} + J_{\mathrm{pen}}H_{\mathrm{pen}}
+ H_{\mathrm{out}}.$$ The coefficients $J_{\mathrm{in}}$,
$J_{\mathrm{prop}}$, and $J_{\mathrm{pen}}$ will later be chosen to be
some polynomials in $n$.  For the term $H_{\mathrm{prop}}$, any valid
history state (a uniform superposition of legal configurations whose
qubit content comes corresponds to the computation with the gates from $\tilde{V}_x$)
will be a zero-energy state. The term $H_{\mathrm{in}}$ raises the
energy of states which do not have ancilla qubits initialized to 0,
which is required in the circuit $\tilde{V}_x$. The role of
$H_{\mathrm{out}}$ is to raise the energy of the states which encode
computations that are not accepted.  Finally, the terms in
$H_{\mathrm{pen}}$ penalize (i.e.\ raise the energy of) locally detectable
illegal configurations which do not have the proper form as described
by equations \eqref{properform1}-\eqref{legalqubitsend} in
Section~\ref{sec:legal}.

We start with the ancilla-checking term $H_{\mathrm{in}}$, defined as
$$
	H_{\mathrm{in}} :=\kets{  \gate^{(1)}} \bras{\gate^{(1)}}_{1}+
	\sum_{i=2}^{n-m}  \kets{  \qubit^{(1)}} \bras{\qubit^{(1)}}_{2i-1}.
$$
By raising the energy of states with qubit content $\ket{1}$, it ensures that in a low-energy state candidate
the ancilla qubits (the first $n-m$) are all 
initially (in the initial configuration \eqref{initialconfiguration} they are located at odd positions in the first block) in the $\ket{0}$ state. Without loss of generality, we assume that the first
round of $V'_x$ consists of identity gates.  This is necessary when we want the
ancilla qubits to remain unpenalized by $H_{\mathrm{in}}$ until they are
moved from the first block into the second block.

The term $H_{\mathrm{out}} :=
\kets{\gate^{(0)}}\bras{\gate^{(0)}}_{2nR}$ checks that when the
computation finishes (the $\gate$ state appears at the very right end
of the qudit chain), the qubit content of the output qubit is
$\ket{1}$. For computations that do not accept, the output qubit state is $\ket{0}$ and $H_{\mathrm{out}}$ penalizes this.

In defining the remaining terms of the Hamiltonian, we will need to be able 
to distinguish between different kinds of configurations of the chain. 
We classify three types: 
\begin{enumerate}
\item {\em legal} configurations are defined in Section~\ref{sec:legal}
  to be the configurations $C_0,\dots, C_K$ occurring during the
  computation with a circuit $\tilde{V}_x$ when starting with the initial configuration
  \eqref{initialconfiguration} with $n$ qubits. All other
  configurations are \emph{illegal}.
  
  \item  {\em locally detectable illegal} configurations are those that contain a pair of neighboring qudits labeled by a pair of symbols that does not occur in the legal configurations. These can be identified and penalized locally by means of a projector onto such a pair.
	 
	\item {\em locally undetectable illegal} configurations are those that are not detectable by local projections,
		but are still not legal, as they do not appear in the legal progression of a computation.
		As shown in Lemma~\ref{lemma:cl}, these states have too many or too few qubits, or an improperly aligned [{\em qubits}] block.
\end{enumerate}

\subsection{The penalty Hamiltonian}
\label{sec:penalty}

The role of $H_{\mathrm{pen}}$ is to ensure that there are no locally
detectable illegal configurations in the computation.  That is, we
wish to leave the legal states unpenalized while raising the energy of the locally detectable illegal ones by
projecting on neighboring pairs of symbols that do not occur in a proper
course of computation described in Section~\ref{sec:circuit}. We call such pairs \textit{forbidden}. Since we have 6 different symbols in the
construction, there are 36 possible neighboring pairs. Furthermore, we can distinguish 5 types of location pairs depending on the parity of the positions and their position with respect to the block boundaries $\bdry$ as listed on the right in Table~\ref{table:legalpairs}. 
We list the 56 allowed pairs of symbols in Table~\ref{table:legalpairs}, which gives us $36\times 5 - 56 = 124$ types of projector terms 
\begin{align}
	\kets{XY}\bras{XY}_{i,i+1},
	\label{projectillegal}
\end{align}
where $XY\in \{\dead, \blnk, \insi, \lmove, \qubit, \gate\}^{\otimes 2}$ is a forbidden pair at a location $(i,i+1)$.
For example, the forbidden pair $\dead \blnk$ (disallowed in all 5 types of locations) is energetically penalized by 
\begin{align}
	\sum_{f=1}^{5} H_{\textrm{pen},f} = \sum_{i=1}^{2nR-1} \ii_{1,\dots,i-1} \otimes 
	 \kets{\dead\blnk} \bras{\dead\blnk}_{i,i+1}
	\otimes \ii_{i+2,\dots,2nR},
\end{align}
while the pair $\qubit \qubit$ is forbidden on even-parity sites (type A,C,E), and is penalized by
\begin{align}
	\sum_{f=93}^{95}H_{\textrm{pen},f} = \sum_{i=1}^{nR}  \ii_{1,\dots,2i-2} \otimes 
	\kets{\qubit \qubit}\bras{\qubit \qubit}_{2i-1,2i}
	\otimes \ii_{2i+1,\dots,2nR}.
\end{align}
To take the qubit content of the $\qubit$ particles into account, as in~\cite{AGIK09} we use the notation
$|A\rangle\langle B| := \sum_s |a_s\rangle\langle b_s|$, meaning that
subspace $B$ is mapped to subspace $A$ in some prescribed way
specified by the pairing of the basis vectors.  Thus, $\kets{\qubit\qubit}\bras{\qubit\qubit}$ preserves the qubit contents as
$\kets{\qubit\qubit}\bras{\qubit\qubit} := \sum_{s,t=0}^1
|\qubit^{(s)}\qubit^{(t)}\rangle\langle \qubit^{(s)}\qubit^{(t)}|$.

Furthermore, to rule out configurations without any qubit-holding particles, we need to penalize the symbols $\{\blnk,\qubit,\lmove,\insi\}$ at the leftmost qudit (only $\dead$ or $\gate$ can appear there) and project onto $\{\dead,\qubit,\lmove,\insi\}$ on the rightmost qudit (only $\gate$ or $\blnk$ are allowed at the right end).
Together, the Hamiltonian  
imposing an energy penalty on configurations containing any
of the forbidden pairs is
\begin{align}
	H_{\textrm{pen}} = \sum_{f=1}^{124} H_{\textrm{pen},f} +H_{\textrm{left}}+H_{\textrm{right}}. \label{Hpen}
\end{align}

Observe that $H_{\mathrm{pen}}$ only catches illegal configurations with
\emph{locally detectable} errors, and there exist illegal configurations
that are not locally detectable, i.e., that have zero energy under
$H_{\mathrm{pen}}$, such as this one with too many qubits
\begin{align}
   \bdry \dead \dead \parity  \qubit \insi \parity  \qubit \insi\bdry  \qubit  \insi\parity  \qubit \blnk\parity \blnk \blnk \bdry .
\end{align}  
To identify these states as illegal, we will have to show they propagate into states with forbidden pairs. First, though,
we want to ensure this propagation, which is the topic of the next section.

\begin{table}
\begin{center}
\begin{adjustbox}{width={\textwidth},totalheight={\textheight},keepaspectratio}
\begin{tabular}{|c|c|c|c|c|c|c|}
\hline
\kern+1bp
 \setlength{\unitlength}{15bp}
\begin{picture}(1,1)
\put(-0.7,0.95){\line(2,-1){2.4}}
\put(-0.4,-0.1){X}
\put(1,0.3){Y}
\end{picture}
\kern+1bp
		& \dead & \blnk & \insi &\lmove &\qubit &\gate \\
\hline
 \dead & 	\ok	&	\xx	&	\xx	& ACE	& ABCE 	&	CD	\\
\hline
 \blnk & 	\xx	&	\ok	&	\xx	&	\xx	&	\xx	&	\xx	\\
\hline
 \insi & 	\xx	&	\xx	&	\xx	& ACE	&   \ok	& AE	\\
\hline
 \lmove & 	\xx	&   ACE	& ACE	&	\xx	& BD	&	\xx	\\
\hline
 \qubit& 	\xx	&   ABCE	&	\ok	& BD	&	BD	&	B	\\
\hline
 \gate & 	\xx	&	DE	&	AC	&	\xx	&	B	&	\xx	\\
\hline
\end{tabular}

\quad 
\begin{tabular}{ll|l}
A: & \preve &$i=2(k-1)n+2j'+1$\\
B: & \, \prodd  &$i=2(k-1)n+2j$\\
C: & \prbleft  &$i=2(k-1)n+1$\\
D: & \, \prbcent &$i=2k'n$\\
E: & \prbrigh  &$i=2kn-1$\\
\hline
&& $1\leq k \leq R$ \\
&& $1\leq k' \leq R-1$ \\
&& $1\leq j \leq n-1$ \\
&& $1\leq j' \leq n-2$ \\
\end{tabular}
\end{adjustbox}
\end{center}
\vspace{5pt}
\caption{The 56 allowed pairs $XY$ of symbols at positions $(i,i+1)$ in the $d=8$ construction according to $H_{\textrm{pen}}$. 
There are 5 types of locations (A, B, C, D, E) for the pair, according to location parity and block-boundary position. For each of the 36 symbol combinations, we list its allowed location types. The forbidden pairs implied by this table are penalized by $H_{\textrm{pen}}$ \eqref{Hpen}.}
\label{table:legalpairs}
\end{table}


\subsection{The propagation Hamiltonian}
We want to check whether the computation on the line of qudits proceeds correctly, in a linear sequence of configurations
$C_0 \leftrightarrow \cdots \leftrightarrow C_t \leftrightarrow
C_{t+1} \leftrightarrow C_{t+2} \leftrightarrow \cdots \leftrightarrow
C_K$ (see Section~\ref{sec:circuit} and the example in Table~\ref{example}),
ensuring the intended unitary operations are applied in the correct order.
The propagation-checking Hamiltonian $H_{\mathrm{prop}}$ should have a low energy only for a state
which is a superposition of all the legal configurations,
with the gates applied to their qubit content as planned.

For now, let us look only at the states from the span of the legal configurations, 
where we want $H_{\mathrm{prop}}$ to give an energy 
penalty to all states except the history states corresponding to the circuit $\tilde{V}_x$.
We would like to construct it as $H_{\mathrm{prop}} =
\sum_{t=0}^{K-1} H_{t}$, where $H_t$ checks the transition from the state $\ket{\psi_{t}}$ to $\ket{\psi_{t+1}}$. For a candidate low-energy state that has a nonzero overlap with $\ket{\psi_{t}}$, it should insist that it has to have the {\em same} amplitude as the state $\ket{\psi_{t+1}}$. 
Any two legal configurations are orthogonal, so if locality did not matter, it would suffice to use projections onto the states $\ket{\psi_t}$ as in \eqref{projectorHprop} in Section~\ref{sec:bg}. However, we want our Hamiltonian to be 2-local. 

The computation of the circuit $\tilde{V}_x$ on an initial state runs according to the
rules in Table~\ref{table:rules}, which are (up to) 4-local.  
A rule $LNOR \leftrightarrow LPQR$ applied at some location corresponds to a transition between states $\ket{\psi_t} = \ket{\cdots LNOR \cdots}$ and $\ket{\psi_{t+1}} =  \ket{\cdots LPQR \cdots}$. In the language of Hamiltonians, this transition is facilitated by  
\begin{align}
	\big(\ii \otimes \big( \ket{LPQR}\bra{LNOR} +  \ket{LNOR}\bra{LPQR}\big) \otimes \ii\big)  \ket{\psi_t} = \ket{\psi_{t+1}} \nonumber
\end{align} 
To penalize states whose overlap with the states $\ket{\psi_t}$ and $\ket{\psi_{t+1}}$ is not the same, we would use
\begin{align}
	& \ii  \otimes \big( \ket{LNOR}\bra{LNOR} + \ket{LPQR}\bra{LPQR}\big) \otimes \ii \nonumber\\
	- & \ii \otimes \big( \ket{LPQR}\bra{LNOR} + \ket{LNOR}\bra{LPQR} \big) \otimes \ii,
	\nonumber
\end{align}
which within the subspace spanned by $\ket{\psi_t}$ and $\ket{\psi_{t+1}}$ projects\footnote{Up to a constant, as it equals $2\ket{\alpha}\bra{\alpha}$ with $\ket{\alpha} = \frac{1}{\sqrt{2}}\left(\ket{\psi_t}+\ket{\psi_{t+1}}\right)$.} onto a state proportional to $\ket{\psi_{t}}-\ket{\psi_{t+1}}$. The equal superposition of the two states is thus an eigenvector with eigenvalue 0. However, we want to use 2-local, not 4-local operators.
If we simply involved only the two particles that actually change ($NO \leftrightarrow PQ$), 
it would be possible that the resulting terms like $\ket{PQ}\bra{NO}$ would apply to several places in a given configuration,
leading to a branching of the legal configuration sequence, 
instead of producing a simple connected line $C_0 \leftrightarrow
\cdots \leftrightarrow C_t \leftrightarrow C_{t+1} \leftrightarrow
\cdots \leftrightarrow C_K$.
This could doom the construction by giving some energy to history states.
However, we will now show how to construct $H_{t}$ from several $2$-qudit terms that
can ``pick out'' and ``check'' the intended transitions between the configurations $C_t$ and $C_{t+1}$. 
The trick involves 2-local terms on surrounding qudits as well.


Let us look at a forward application of a rule that changes the qudits $(i,i+1)$, 
taking them from a sub-configuration $NO_{(i,i+1)}$ to a sub-configuration $PQ_{(i,i+1)}$. 
We constructed the legal configurations (see Table~\ref{table:evolve}) so that this rule is applicable only to a configuration
that is uniquely identifiable by a sub-configuration $XY_{(j,j+1)}$ on some nearby qudits $(j,j+1)$.
Similarly, the backwards applicability of this rule is uniquely identifiable by a sub-configuration $ZW_{(k,k+1)}$
on some nearby qudits $(k,k+1)$.  
We now write a Hamiltonian checking the application of this rule as
\begin{align}
	H_{\textrm{prop}, i}^{(\textrm{rule})} &= \ket{XY}\bra{XY}_{j,j+1} + \ket{ZW}\bra{ZW}_{k,k+1} - \ket{PQ}\bra{NO}_{i,i+1} - \ket{NO}\bra{PQ}_{i,i+1}.
	\label{Hpropterm}
\end{align}
In the simplest case, $XY=NO$, $ZW=PQ$ and $i=j=k$, so that $XY, ZW, NO, PQ$ all involve the same pair of particles $(i,i+1)$. 
For a more complicated case, let us look at rule 4(b) $\qubit \parity \blnk \blnk \longleftrightarrow \qubit \parity \lmove \blnk$ from Table~\ref{table:rules}. The forward applicability of the rule is uniquely identified by the substring $\qubit \parity \blnk$ on the first two particles, while the backwards applicability of this rule is uniquely identified by the substring $\lmove \blnk$ on the second and third particles. The Hamiltonian term checking this rule will be given in \eqref{Hmake}.

Let us now look at an example from a unary clock construction, to see that history states retain a zero-energy for 
a Hamiltonian of the type \eqref{Hpropterm}.


\subsubsection{Analogy with [KKR06]}

As an example, we recall \cite{KKR06}, where the propagation
Hamiltonian was reduced from 3-local to 2-local. 
There, checking the progress of a unary clock register
$\ket{s}=\kets{1\cdots 1_s 0\cdots 0}$ can be done with a 3-local Hamiltonian
\begin{align}
	H_{t} =	\big( 
	\kets{100}\bras{100}
	+	\kets{110}\bras{110}
	-	\kets{110}\bras{100}
	-	\kets{100}\bras{110}
	\big)_{t,t+1,t+2},
\end{align}
uniquely picking the states $\ket{t}$ and $\ket{t+1}$ for which the transition rule $100 \leftrightarrow 110$ applies. 
Instead, we can make it out of 2-local (and 1-local) terms 
\begin{align}
H'_{t} = \kets{10}\bras{10}_{t,t+1} 
+	\kets{10}\bras{10}_{t+1,t+2}
-	\kets{1}\bras{0}_{t+1}
-	\kets{0}\bras{1}_{t+1}.
	\label{KKRprop}
\end{align}
with the first two terms uniquely identifying the places where the rule should apply, while the last two (transition) terms are ambiguous in their applicability. 
The price for the decrease in locality are ``mistimed''
transitions such as
$\ket{11\mathbf{1}100}\rightarrow\ket{11\mathbf{0}100}$ in the unary clock
register. However, observe that the expectation value of this Hamiltonian in the uniform
superposition of unary clock states is zero, i.e. 
\[
	\frac{1}{T+1}\sum_{r,s=0}^{T} \bra{r} H'_{t}  \kets{s} = 0,
\]
because the mistimed transitions in $H'_{t}$ take the state out of
the legal subspace, to states orthogonal to any of the proper unary
clock states. On the other hand, the transitions taking place at
proper places are easily shown to result in 0 energy. Thus, the
restriction of $H'_{i}$ to the legal clock subspace spanned by
$\{\ket{0},\ket{1},\dots,\ket{T}\}$ works exactly as the Hamiltonian
$H^{\textrm{unary}}_{\textrm{prop},i} = \ket{t}\bra{t} +
\ket{t+1}\bra{t+1} - \ket{t+1}\bra{t}- \ket{t+1}\bra{t}$ from
\eqref{projectorHprop}. We conclude that a correct history state (a
superposition of all legal states) has expectation energy zero under
the decreased-locality propagation-checking Hamiltonian
\eqref{KKRprop} from \cite{KKR06}. 

The projector terms in $H'_t$ thus picked the applicability place uniquely,
while the mistimed transitions coming from the last two terms
took the state out of the legal subspace (to non-unary clock states).
We will now use this insight to construct our $H_{\textrm{prop}}$ from 2-local terms.
However, our task is more complicated, because the unary-clock propagation rule $100 \rightarrow 110$ 
for a certain location applied just once in a sequence of proper clock states. In our case,
a rule for moving a qubit (or a gate particle) at a given location in the chain could connect configurations
$C_t \leftrightarrow C_{t+1}$ as well as some other configurations $C_{t'} \leftrightarrow C_{t'+1}$
with the data in the chain shifted by a few positions.


\subsubsection{Explicit form of the propagation-checking terms}
Instead of writing the propagation Hamiltonian as a sum of terms
$H_t$, we choose to write it out as a sum of terms $H^{(\textrm{rule
  }\rho)}_{\textrm{prop},i}$ corresponding to different rules $\rho$
applied at location pairs $(i,i+1)_\rho$ wherever rule $\rho$ is
applicable. This generalization is required because one rule $\rho$ for a pair of particles
can facilitate legal transitions between several configuration pairs.
The propagation Hamiltonian is then  
\begin{align}
	H_{\textrm{prop}} = \sum_{\rho =1}^{6} \sum_{(i,i+1)_{\rho}} H^{(\textrm{rule }\rho)}_{\textrm{prop},i}. \label{Hprop}
\end{align}
and we want its application to a state $\ket{\psi_t}$ (corresponding to a legal configuration $C_t$ with $2\leq t \leq K$) to result in
\begin{align}
	H_{\textrm{prop}} \ket{\psi_t} = 2 \ket{\psi_t} - \ket{\psi_{t-1}} - \ket{\psi_{t+1}} + \textrm{illegal but locally detectable states}. \label{property}
\end{align}

The propagation term corresponding to rule \ref{rule:gate} ($\gate \parity \qubit \longleftrightarrow
  U_m(\qubit \parity \gate) $) in Table~\ref{table:rules} is simple, as it involves only the sites $i$ and $i+1$ and does not create bad transitions. We want to check the transfer of $\gate$ to the right and the application of the gate $U_{t_i}$ (corresponding to the location $i$) to the qubit content of the two neighboring sites. This is done by 
\begin{align}
	H_{\textrm{prop},i}^{(\textrm{rule }\ref{rule:gate})} &=
	\phantom{U_{t_i}} \kets{\gate\qubit}\bras{\gate\qubit}_{i,i+1} 	
	+ \phantom{U_{t_i}} \kets{\qubit\gate}\bras{\qubit\gate}_{i,i+1}  \label{Hgate}\\
	&-  U_{t_i} \kets{\qubit\gate}\bras{\gate\qubit}_{i,i+1} 
	- U_{t_i}^{\dagger} \kets{\gate\qubit}\bras{\qubit\gate}_{i,i+1},
\end{align}
and this term appears only for locations $(i,i+1)$ of the type B in Table~\ref{table:legalpairs}.

We continue with rule \ref{rule:move} ($ \parity \gate \insi \parity  \longleftrightarrow \parity  \insi \gate \parity $, $ \bdry \gate \insi \parity  \longleftrightarrow \bdry \dead \gate \parity  $, $\parity  \gate \blnk \bdry \longleftrightarrow  \parity  \insi \gate \bdry $) for moving the $\gate$ from position $i$ to $i+1$.
Depending on the location in the chain, the Hamiltonian term reads
\begin{align}
	H_{\textrm{prop},i}^{(\textrm{rule }\ref{rule:move})} &=
	\kets{\gate O}\bras{\gate O}_{i,i+1} 
	+ \kets{P\gate}\bras{P\gate}_{i,i+1} \nonumber\\
	&-   \kets{P\gate}\bras{\gate O}_{i,i+1} 
	-   \kets{\gate O}\bras{P\gate}_{i,i+1}, 
		\label{Hmove}
\end{align}
with $PO=\insi\insi$ for locations $(i,i+1)$ of type A,
$PO=\dead\insi$ for locations of type C, and
$PO=\insi\blnk$ for locations of type E.

The propagation terms for rule \ref{rule:make} ($\gate \bdry \blnk \blnk \longleftrightarrow \qubit \bdry \lmove \blnk$, $\qubit \parity \blnk \blnk \longleftrightarrow \qubit \parity \lmove \blnk$) govern the creation of the symbol $\lmove$. We now involve three particles, but again, only two at a time,
leaving the Hamiltonian 2-local:
\begin{align}
	H_{\textrm{prop},i}^{(\textrm{rule }\ref{rule:make})} &=
	\kets{X\blnk}\bras{X\blnk}_{i,i+1} 
	+ \kets{\lmove\blnk}\bras{\lmove\blnk}_{i+1,i+2} \nonumber\\
	&-  \kets{\qubit \lmove}\bras{X\blnk}_{i,i+1} 
	-   \kets{X\blnk}\bras{\qubit \lmove}_{i,i+1} 
		\label{Hmake}
\end{align}
with $X=\gate$ at locations of the type D 
and $X=\qubit$ at locations of the type B. 
Only the projector term identifying the backwards applicability of rule \ref{rule:make}
 involves a particle pair different from $(i,i+1)$.

Rule \ref{rule:left} ($\ppp \qubit \bothp \lmove \longleftrightarrow \ppp\lmove \bothp \qubit$, $\bothp \insi  \lmove \bothp \longleftrightarrow \bothp \lmove \insi\bothp$) pushes $\lmove$ to the left, and its checking Hamiltonian is again simple:
\begin{align}
	H_{\textrm{prop},i}^{(\textrm{rule }\ref{rule:left})} &=
	 \kets{X\lmove}\bras{X\lmove}_{i,i+1} 
	 + \kets{\lmove X}\bras{\lmove X}_{i,i+1} \nonumber\\
	&-     \kets{\lmove X}\bras{X\lmove}_{i,i+1}
	- \kets{X\lmove}\bras{\lmove X}_{i,i+1} 
	\label{Hleft}
\end{align}
with $X=\insi$  at locations of the type ACE and 
with $X=\qubit$  at locations of the type BD.

The Hamiltonian for rule \ref{rule:kill} ($\dead \lmove \bdry \qubit \longleftrightarrow \dead \dead \bdry \gate$, $\dead \lmove \parity \qubit \longleftrightarrow \dead
  \dead \parity \qubit$) kills the symbol $\lmove$ and
mirrors the ones for rule \ref{rule:make}:
\begin{align}
	H_{\textrm{prop},i}^{(\textrm{rule }\ref{rule:kill})} &=
	\kets{\dead \lmove}\bras{\dead \lmove}_{i-1,i} 
	+ \kets{\dead W}\bras{\dead W}_{i,i+1} \nonumber\\
		&-  \kets{\dead W}\bras{\lmove \qubit}_{i,i+1} 
	-   \kets{\lmove \qubit}\bras{\dead W}_{i,i+1} 
	\label{Hkill}
\end{align}
with $W=\gate$ at locations of the type D 
and $W=\qubit$ at locations of the type B. 


Rule \ref{rule:moveq} is the most complicated one
since its definition is 4-local. We reduce the locality by
looking only at qudit pairs $(i-1,i)$ and $(i+1,i+2)$ to identify 
the applicability of this rule to states $\ket{\psi_t}$ and $\ket{\psi_{t+1}}$
which can be connected by an application of rule~\ref{rule:moveq}.  Note
that the pair $(i,i+1)$ has to be of the type ACE.

We begin by writing out the propagation terms corresponding to each of
the possible transitions in rule \ref{rule:moveq}. Rule
\ref{rule:moveq1} ($ \qubit \bothp \qubit \insi \bothp\qubit   
		\longleftrightarrow \qubit \bothp \insi \qubit \bothp\qubit  $) applies to pairs $(i,i+1)$ of type $A$, $C$, and
$E$: 
\begin{align}
H_{\textrm{prop},i}^{(\textrm{rule }\ref{rule:moveq1})} &=  \kets{\qubit\qubit}\bras{\qubit\qubit}_{i-1,i}
	+	\kets{\qubit\qubit}\bras{\qubit\qubit}_{i+1,i+2} \nonumber \\
  & - \kets{\insi\qubit}\bras{\qubit\insi}_{i,i+1}
	- 	\kets{\qubit\insi}\bras{\insi\qubit}_{i,i+1}.
\end{align}
Rule \ref{rule:moveq2} ($ \dead \parity \qubit \insi \bothp\qubit   \longleftrightarrow \dead \parity \dead \qubit \bothp \qubit  $) applies to pairs $(i,i+1)$ of type A and E:
\begin{align}
  H_{\textrm{prop},i}^{(\textrm{rule }\ref{rule:moveq2})}   &= \kets{\dead\qubit}\bras{\dead\qubit}_{i-1,i} + \kets{\qubit\qubit}\bras{\qubit\qubit}_{i+1,i+2} \nonumber \\
  & - \kets{\dead\qubit}\bras{\qubit\insi}_{i,i+1}
	- 	\kets{\qubit\insi}\bras{\dead\qubit}_{i,i+1}. 
\end{align}
Rule \ref{rule:moveq3} ($ \qubit \bothp \qubit \blnk \parity \blnk   \longleftrightarrow \qubit \bothp \insi \qubit \parity \blnk  $) acts on pairs of type A and C:
\begin{align}
H_{\textrm{prop},i}^{(\textrm{rule }\ref{rule:moveq3})} &= \kets{\qubit\qubit}\bras{\qubit\qubit}_{i-1,i} + \kets{\qubit\blnk}\bras{\qubit\blnk}_{i+1,i+2} \nonumber \\ 
  & - \kets{\insi\qubit}\bras{\qubit\blnk}_{i,i+1}
	- 	\kets{\qubit\blnk}\bras{\insi\qubit}_{i,i+1}.
\end{align}
Finally, rule \ref{rule:moveq4} ($ \dead \bothp \qubit \blnk \bothp \blnk   \longleftrightarrow \dead \bothp \dead \qubit \bothp \blnk  $) handles a special type of illegal configuration
that contains only a single qubit-holding particle. In combination with rules \ref{rule:make}-\ref{rule:kill}, it helps to move this qubit until it reaches a
locally-detectable illegal configuration. For locations $(i,i+1)$ of type A, C and E, we write
\begin{align}
H_{\textrm{prop},i}^{(\textrm{rule }\ref{rule:moveq4})} &= \kets{\dead\qubit}\bras{\dead\qubit}_{i-1,i} + \kets{\qubit\blnk}\bras{\qubit\blnk}_{i+1,i+2} \nonumber \\
  & - \kets{\dead\qubit}\bras{\qubit\blnk}_{i,i+1}
	- 	\kets{\qubit\blnk}\bras{\dead\qubit}_{i,i+1}  .
\end{align}

To obtain the overall Hamiltonian for rule \ref{rule:moveq}, we do not simply sum these four terms,
as we would include the terms $\kets{\dead\qubit}\bras{\dead\qubit}_{i-1,i}$,
$\kets{\qubit\blnk}\bras{\qubit\blnk}_{i+1,i+2}$,
$\kets{\qubit\qubit}\bras{\qubit\qubit}_{i-1,i}$,
$\kets{\qubit\qubit}\bras{\qubit\qubit}_{i+1,i+2}$ twice. 
The function of these projectors is to pick out the `before'
and `after' configurations of the corresponding transition rule. Since
we want each legal configuration to picked out exactly once, we
include them only in a single copy, i.e.  
\begin{align}
	H_{\textrm{prop},i}^{(\textrm{rule }\ref{rule:moveq})} &=
	\kets{\qubit\qubit}\bras{\qubit\qubit}_{i-1,i}
	+ 	\kets{\qubit\qubit}\bras{\qubit\qubit}_{i+1,i+2}
		\label{Hmoveq}
		\\
	&+
	\kets{\dead\qubit}\bras{\dead\qubit}_{i-1,i}
	+ 	\kets{\qubit\blnk}\bras{\qubit\blnk}_{i+1,i+2} \label{Hmoveq2}\\
	&-
	\kets{\insi\qubit}\bras{\qubit\insi}_{i,i+1}
	- 	\kets{\qubit\insi}\bras{\insi\qubit}_{i,i+1} \label{Hmoveq3}\\
	&-
	\kets{\dead\qubit}\bras{\qubit\blnk}_{i,i+1}
	- 	\kets{\qubit\blnk}\bras{\dead\qubit}_{i,i+1} \label{Hmoveq4}\\
	&-
	\kets{\dead\qubit}\bras{\qubit\insi}_{i,i+1}
	- 	\kets{\qubit\insi}\bras{\dead\qubit}_{i,i+1} \quad \textrm{only at location types AE} \label{Hmoveq5}\\
	&-
	\kets{\insi\qubit}\bras{\qubit\blnk}_{i,i+1}
	- 	\kets{\qubit\blnk}\bras{\insi\qubit}_{i,i+1} \quad \textrm{only at location types AC} \label{Hmoveq6}
\end{align}
where the first four lines apply at locations $(i,i+1)$ of the types ACE, and the  last two lines apply only at the location types listed (AE and AC). The projector terms in this Hamiltonian term applied to a legal configuration now gives something nonzero 
only when rule~\ref{rule:moveq} could be applied to this legal configuration -- and induces exactly {\em one} forward and {\em one} backward legal transition. Possibly, it could also induce illegal transitions, which lead to illegal states detectable by $H_{\textrm{pen}}$). On the other hand, when rule~$\ref{rule:moveq}$ (in its 4-local form) is not applicable to a given configuration, we get no projection terms, only transitions to illegal states.

Note that we need to fix the prescriptions at the ends of the chain. We do this by omitting the particles with positions with $i-1<1$, $i+1>2nR$ or $i+2>2nR$, e.g., using only a single-particle projector $\kets{\qubit}\bras{\qubit}_1$ as the first term in \eqref{Hmoveq} at the location pair (1,2). Together, all these terms make up $H_{\textrm{prop}}$.


\subsubsection{Applying the propagation Hamiltonian: examples}
Let us see the Hamiltonian for rule~\ref{rule:moveq} in action. We list a few examples, applying it in a location of the type E (with a block boundary on the right), first to legal configurations:
\begin{align*}
\begin{adjustbox}{width={\textwidth},totalheight={\textheight},keepaspectratio}
	\begin{tabular}{|c|l|}
		\hline
			a legal configuration  & applying $H^{(\textrm{rule }\ref{rule:moveq})}_{\textrm{prop},i}$ for $(i,i+1)=(5,6)$ gives\\
		\hline
			$ C_1=\dead \dead \parity \dead  \qubit\parity \qubit \insi \bdry \qubit\blnk \parity \blnk \blnk \mgdots$&
				$+ \dead \dead \parity \dead  \qubit\parity \qubit \insi \bdry \qubit\blnk \parity \blnk \blnk \mgdots$ (projection)\\
				&
				$- \dead \dead \parity \dead  \qubit\parity \insi \qubit \bdry \qubit\blnk \parity \blnk \blnk \mgdots$ (legal transition)\\
				&
				$- \dead \dead \parity \dead  \qubit\parity \dead \qubit \bdry \qubit\blnk \parity \blnk \blnk \mgdots$ (locally detectable)\\
		\hline
			$ C_2=\dead \dead \parity \dead  \dead \parity \qubit \insi \bdry \qubit\insi \parity \qubit\blnk \mgdots$&
				$+ \dead \dead \parity \dead  \dead \parity \qubit \insi \bdry \qubit\insi \parity \qubit\blnk \mgdots$ (projection)	\\
				&
				$- \dead \dead \parity \dead  \dead \parity \dead \qubit \bdry \qubit\insi \parity \qubit\blnk \mgdots$ (legal transition)	\\
				&
				$- \dead \dead \parity \dead  \dead \parity \insi \qubit \bdry \qubit\insi \parity \qubit\blnk \mgdots$ (locally detectable)	\\
		\hline
			$ C_3=\dead \qubit\parity \lmove \insi \parity \qubit \insi \bdry \qubit\blnk \parity \blnk \blnk \mgdots$&
				$- \dead \qubit\parity \lmove \insi \parity \insi \qubit \bdry \qubit\blnk \parity \blnk \blnk \mgdots$ (locally detectable)\\
				&
				$- \dead \qubit\parity \lmove \insi \parity \dead \qubit \bdry \qubit\blnk \parity \blnk \blnk \mgdots$ (locally detectable)\\
		\hline
	\end{tabular}
  \end{adjustbox}
\end{align*}
For the first configuration $C_1$, the pair (5,6) is indeed where rule \ref{rule:moveq} should apply. Thus we get a projection from \eqref{Hmoveq}, and a legal transition from \eqref{Hmoveq3}. There is an additional illegal transition from \eqref{Hmoveq5}, locally detectable by the illegal pair $\qubit \dead$.
The second configuration $C_2$ should transform forward by applying rule \ref{rule:moveq}. The configuration is projected by \eqref{Hmoveq2}, has a legal transition from \eqref{Hmoveq5} 
and an extra illegal one from \eqref{Hmoveq3} with the bad substring $\dead \insi$.
For the third configuration $C_3$, rule~\ref{rule:moveq} should not apply (the proper-transition producing rule is now rule~\ref{rule:left}, involving $\lmove$) -- thus we get nothing from the projection terms \eqref{Hmoveq}-\eqref{Hmoveq2}. Instead, we get two transitions to illegal states -- \eqref{Hmoveq3} creates a state with a bad substring $\insi \insi$ and \eqref{Hmoveq5} makes a state detectable by the pair $\insi \dead$. 

Let us look at one more example, checking what  $H^{(\textrm{rule }\ref{rule:moveq})}_{\textrm{prop},i}$ does to an illegal, allowed but not locally detectable configuration. This special case is crucial for detecting configurations with badly aligned blocks or with too few/too many qubits.
\begin{align}
\begin{adjustbox}{width={\textwidth},totalheight={\textheight},keepaspectratio}
	\begin{tabular}{|c|l|}
		\hline
			an allowed but illegal configuration  & applying $H^{(\textrm{rule }\ref{rule:moveq})}_{\textrm{prop},i}$ for the middle pair\\
		\hline
			$ C_4=\mgdots \dead  \dead \parity \qubit \blnk \bdry \blnk \blnk  \mgdots$&
				$ + \dead \dead \parity \dead  \dead \parity \qubit \blnk \bdry \blnk \blnk \parity \blnk\blnk \mgdots$ (projection)\\
				&
				$ - \dead \dead \parity \dead  \dead \parity \dead \qubit \bdry \blnk \blnk \parity \blnk\blnk \mgdots$ (loc. detectable)\\
		\hline
	\end{tabular}
  \end{adjustbox}
	\label{allowedillegal}
\end{align}
The configuration $C_4$ is projected once by \eqref{Hmoveq2} and an illegal transition from \eqref{Hmoveq4} makes a configuration with a  forbidden pair $\qubit\bdry\blnk$. Note that we did not obtain a legal transition, even though rule~\ref{rule:moveq} was applicable and gave us a projection term. In Section \ref{locallyundetectablesection} we will translate this into a lower bound on the ground state energy for such states.


\subsubsection{Classifying the legal and illegal transitions}
When $H_{\textrm{prop}}$ \eqref{Hprop} acts on a state of the system, it induces
changes in the configuration (besides sometimes performing a two-qubit unitary
operation). This construction differs from the one in~\cite{AGIK09}
in that the changes can occur at more than one location along the chain. 
This is readily apparent when one considers the action of the
whole propagation Hamiltonian on the 
state with a configuration such as
$\mgdots \dead \dead \bdry \gate \insi \parity \qubit \insi \parity \qubit \blnk \bdry \blnk \blnk \mgdots$ (the first line in Table~\ref{example}).
We obtain
\begin{align}
	H_{\mathrm{prop}} \kets{\mgdots \dead \dead \bdry \gate \insi \parity \qubit \insi \parity \qubit \blnk \bdry \blnk \blnk \mgdots}
	=
	&+
	\kets{\mgdots \dead \dead \bdry \gate \insi \parity \qubit \insi \parity \qubit \blnk \bdry \blnk \blnk \mgdots} \label{Hexample}\\
	&-
	\kets{\mgdots \dead \dead \bdry \dead \gate \parity \qubit \insi \parity \qubit \blnk \bdry \blnk \blnk \mgdots} \nonumber\\
	&+
	\kets{\mgdots \dead \dead \bdry \gate \insi \parity \qubit \insi \parity \qubit \blnk \bdry \blnk \blnk \mgdots} \nonumber\\
	&-
	\kets{\mgdots \dead \lmove \bdry \qubit \insi \parity \qubit \insi \parity \qubit \blnk \bdry \blnk \blnk \mgdots} \nonumber\\
	&-
	\kets{\mgdots \dead \dead \bdry \gate \insi \parity \insi \qubit \parity \qubit \blnk \bdry \blnk \blnk \mgdots} \nonumber\\
	&-
	\kets{\mgdots \dead \dead \bdry \gate \insi \parity \dead \qubit \parity \qubit \blnk \bdry \blnk \blnk \mgdots} \nonumber\\
	&-
	\kets{\mgdots \dead \dead \bdry \gate \insi \parity \qubit \insi \parity \dead \qubit \bdry \blnk \blnk \mgdots} \nonumber
\end{align}
with the first 2 terms coming from \eqref{Hmove}, connected with  rule \ref{rule:move} for moving the $\gate$ particle, 
the second 2 terms coming from \eqref{Hkill}, connected with backward application of rule \ref{rule:kill} for making $\lmove$ disappear.
These 4 terms are exactly what we would like. However, we also obtain the three transitions to locally detectable states on the 5-7th lines, 
due to \eqref{Hmoveq3}-\eqref{Hmoveq5}, connected with rule \ref{rule:moveq} for moving a qubit $\qubit$.
The way the legal and locally detectable terms are produced by $H_{\mathrm{prop}}$ in our construction obeys certain rules.

We can check  that
$H_{\mathrm{prop}}$ acting on a state $\ket{\psi_t}$ with a \emph{legal}
configuration (i.e. one appearing in a computation as in Table~\ref{example}, with a correct number of qubits, properly aligned between boundaries) produces a superposition
$H_{\mathrm{prop}} \ket{\psi_t}$ that contains
\begin{enumerate}
\item The original legal state with amplitude 2 (for two rules that apply to it), 
	except for $t=0$ and $t=K$, where the amplitude is 1 (only a single rule applies to those two states).
\item Two new legal configurations: one due to a `forward' transition
	to $\ket{\psi_{t+1}}$ and one due to a `backward' transition to $\ket{\psi_{t-1}}$.
	Note that for $t=0$ and $t=K$ we only get one legal transition.
\item Some illegal terms, which are all locally detectable with $H_{\mathrm{pen}}$ 
	(such as the 5th line in \eqref{Hexample}, with the illegal combination $\insi\insi$).
\end{enumerate}

Points 1 and 2 are a property of our construction with projector terms uniquely picking out only the ``active'' spots in a given configuration. We discuss the verification of point 3 (verifying that transition terms applied at inappropriate times are always locally detectable) in the next section.

Note that there exist allowed (not containing one of the forbidden pairs) configurations that are not locally detectable (such as the configuration $C_4$ in \eqref{allowedillegal} with a single qubit). These are either improperly aligned or have an incorrect number of qubits. For some of these states, $H_{\textrm{prop}}$ gives only one transition to another allowed state, while it still projects onto the state twice -- this will be used to show the energy of such locally undetectable states is still high. For example the configuration from \eqref{allowedillegal} is projected twice by the terms in $H_{\textrm{prop}}$ corresponding to rules \ref{rule:moveq} and (the backwards application of) \ref{rule:kill}.

\vspace*{12pt}
\noindent
{\bf Lemma} \label{lemma:cl}
  A configuration that does not contain any of the forbidden pairs
  (i.e.\ those penalized by $H_{\mathrm{pen}}$) is either one of the legal
  configurations (configurations that occur in the course of a
  computation), or (i) has a [qubits] string of incorrect length, or
  (ii) has a [qubits] string of the right length, but improperly
  aligned with the block boundaries.

\vspace*{12pt}
\noindent

{\bf Proof:}
  Careful checking of the allowed pairs at positions
  $(i,i+1)$ for odd $i$ and even $i$ from Table~\ref{table:legalpairs} 
  implies the allowed joining of symbol pairs given in Table~\ref{table:activity}).
  This in turn restricts the legal/allowed configurations to form (\lxx\,$\cdots$\lxx)$[qubits]$(\luu\,$\cdots$\luu)
  where [{\em qubits}] is a nonzero string with the structure \eqref{properform1}-\eqref{legalqubitsend}.
   
The only configurations that are not ruled out by these considerations
are: (i) the legal configurations, (ii) configurations with a
[\emph{qubits}] string of the wrong length (not containing exactly $n$ qubits), 
(iii) configurations with a [\emph{qubits}] string with the right number of qubits, but improperly aligned with the blocks (e.g. 
$\cdots \dead \parity \dead \qubit \bdry \insi \qubit \parity \gate \insi \parity \qubit \blnk \parity \blnk \cdots$
which eventually evolves to $\cdots \dead \parity \dead \qubit \bdry \insi \qubit \parity \insi \qubit \parity \gate \blnk \parity \blnk \cdots$
with a bad pair $\parity \gate \blnk \parity$ indicative of a misaligned block). $\square$


\section{Completeness}
\label{sec:complete}

Suppose there exists a witness, $\vert \xi \rangle$, that is accepted
by $V_x$ with probability $\geq 1 - \epsilon$. Beginning with the
initial state $\ket{\psi_0}$ \eqref{initialconfiguration} that has
$n$ qubits with qubit content $\ket{0^{n-m}}\otimes\ket{\xi}$, we get the history
state $\vert \eta \rangle = \frac{1}{\sqrt{K+1}}\sum_{t=0}^K \vert
\psi_t \rangle$ associated with circuit $\tilde{V}_x$. The
configurations occurring in this superposition are exactly the legal
configurations. Given that all the ancilla qubits were initially in
the $\ket{0}$ state, $H_{\mathrm{in}}$ evaluates to zero on $\vert \eta
\rangle$. Since all the configurations in the superposition are legal,
$H_{\mathrm{pen}}$ also evaluates to zero.

We next note the following facts about the legal configurations to be
used in the claim.

\vspace*{12pt}
\noindent
\begin{fact} \label{fact1}
	Any legal configuration can contain at most one substring on 
	the lefthand side of the transition rules~\ref{rule:gate}--\ref{rule:kill}. This means that to any legal 
	configuration, at most one of the transition rules can apply in the forward direction. 
	Furthermore, a legal configuration can contain at most one substring $XY_{(j,j+1)}$ 
	(and thus be connected to a single projector term of the type $\ket{XY}\bra{XY}_{j,j+1}$
	in all of $H_{\textrm{prop},i}^{\textrm{rule }\rho}$ \eqref{Hpropterm}).
\end{fact}

\vspace*{12pt}
\noindent
The first part of this fact can be verified by inspection of the list of legal configurations in Table~\ref{example} and the transitions that
can be applied to them in Table~\ref{table:rules}.
To check the second part (about the projector terms), in Table~\ref{table:XYZW} we list the substrings $XY$ identifying active spots in legal configurations from the projector terms of the type $\ket{XY}\bra{XY}_{j,j+1}$ in all of the $H_{\textrm{prop},i}^{\textrm{rule }\rho}$. An inspection of the legal sequence again shows that each state in it has only one spot where one of the substrings $XY$ appears (at a proper location with respect to the boundaries).

\vspace*{12pt}
\noindent
\begin{fact} \label{fact2}
	Any legal configuration can contain at most one substring from the righthand side of the transition
        rules~\ref{rule:gate}--\ref{rule:kill} (i.e. at most one transition rule applies to it in the backward direction).  
        	Furthermore, a legal configuration can contain at most one substring $ZW_{(k,k+1)}$ 
	(and thus be connected to a single projector term of the type $\ket{ZW}\bra{ZW}_{k,k+1}$
	in all of $H_{\textrm{prop},i}^{\textrm{rule }\rho}$ \eqref{Hpropterm}).
\end{fact}

\vspace*{12pt}
\noindent

\vspace*{12pt}
\noindent
\begin{fact} \label{fact4}
	For a legal configuration $C_t$, there can be multiple places containing one of the substrings $NO_{(i,i+1)}$
	or $PQ_{(i,i+1)}$ from all of the terms in $H_{\textrm{prop},i}^{\textrm{rule }\rho}$ \eqref{Hpropterm}.
	However, exchanging any $NO \rightarrow PQ$ in $C_t$ leads to locally detectable illegal configurations for all cases except one, which gives the legal (following) configuration $C_{t+1}$. 
	Similarly, exchanging any $PQ \rightarrow NO$ in $C_t$ leads to locally detectable illegal configurations for all cases except one, which gives the legal (preceding) configuration $C_{t-1}$. 
\end{fact}

\vspace*{12pt}
\noindent
We have chosen the $PQ$'s and $NO$'s so that they work properly where they should, and always produce locally detectable illegal configurations when used at ``wrong times'' (i.e. improper locations). This can be checked by careful inspection of the transition rules and terms in $H_{\textrm{prop},i}^{\textrm{rule }\rho}$.

\begin{table}
\begin{center}
\begin{adjustbox}{width={\textwidth},totalheight={\textheight},keepaspectratio}
\begin{tabular}{|c|l|l|}
\hline
	location type & $XY$ (for a forward transition) & $ZW$ (for a backward transition) \\
\hline
	(A) \preve  & 
			$
			   \parity  \insi \lmove \parity , \parity \dead \lmove \parity
			$ & 
			$
			 \parity \lmove \insi \parity, \parity \lmove \blnk \parity
			$ \\
\hline
	(B) \, \prodd\,  & 
			$
			\gate\parity\qubit, \qubit\parity \qubit, \dead \parity \qubit,  \qubit \parity \blnk, \qubit \parity \lmove
			$ & 
			$
			 \qubit \parity \gate, \qubit\parity\qubit, \dead \parity \qubit,  \qubit \parity \blnk, \lmove \bothp \qubit
			$ \\
\hline
	(C) \prbleft  & 
			$\bdry  \insi \lmove  , \bdry \dead \lmove 
			$ & 
			$\bdry \lmove \insi , \bdry \lmove \blnk, \bdry \dead  \gate
			$ \\
\hline
	(D) \, \prbcent\,  & 
			$\gate \bdry \blnk, \qubit \bdry \lmove
			$ & 
			$\dead \bdry \gate, \lmove \bdry \qubit
			$ \\
\hline
	(E) \prbrigh  & 
			$ \insi \lmove \bdry ,  \dead \lmove \bdry, \gate \blnk \bdry
			$ & 
			$  \lmove \insi \bdry,  \lmove \blnk \bdry
			$ \\
\hline
\end{tabular}
\end{adjustbox}
\end{center}
\label{table:XYZW}
\vspace{5pt}
\caption{Substrings identifying active spots in legal configurations. We list all of the substrings appearing in the projector terms of the type $\ket{XY}\bra{XY}_{j,j+1}$ and $\ket{ZW}\bra{ZW}_{k,k+1}$ from all of the $H_{\textrm{prop},i}^{\textrm{rule }\rho}$ terms \eqref{Hgate}, \eqref{Hmove}, \eqref{Hmoveq}, \eqref{Hmake}, \eqref{Hleft} and \eqref{Hkill}.
Finding a substring XY of a legal configuration $C_t$ uniquely indicates that the configuration is connected to some configuration $C_{t+1}$ ahead of it. Similarly, finding a substring ZW uniquely locates a backward transition to some $C_{t-1}$.
}
\end{table}

\vspace*{12pt}
\noindent
\begin{claim}
  For any history state $|\eta\rangle$ with an initial configuration $C_0$ from \eqref{initialconfiguration},
  $\langle \eta|H_{\mathrm{prop}}|\eta
  \rangle =0$.
\end{claim}

\vspace*{12pt}
\noindent

{\bf Proof:}
 Let us see what happens when we apply $H_{\textrm{prop}}$ to a state $\ket{\psi_t}$ with a legal configuration $C_t$. 
 The propagation Hamiltonian is made from terms of the type \eqref{Hpropterm}, with projection terms built from substrings $XY$ and $ZW$, and transition terms exchanging substrings $NO$ for $PQ$ and vice versa. 

 First, due to Fact \ref{fact1}, a legal configuration $C_t$ contains only one substring $XY_{(j,j+1)}$, and this projection term will apply, producing  $\ket{\psi_t}$. Second, due to Fact \ref{fact4}, the corresponding transition term $NO_{(i,i+1)} \rightarrow PQ_{(i,i+1)}$ will apply, producing the state $-\ket{\psi_{t+1}}$. Possibly, other transition terms for other substrings $NO \rightarrow PQ$ will apply at a different (or the same) location, but these all produce locally detectable configurations, which are orthogonal to legal ones. 
 Third, due to Fact \ref{fact2} we get a single projection term because of the unique substring $ZW_{(k,k+1)}$, producing $\ket{\psi_t}$ again. Fourth, again due to Fact \ref{fact4}, we get a single legal transition to $-\ket{\psi_{t-1}}$, and possible other, illegal, detectable states.
 
In sum, the action of $H_{\textrm{prop}}$ on $\ket{\psi_t}$ (for $1\leq t \leq K-1$) produces, $-\ket{\psi_{t-1}} + 2\ket{\psi_t} - \ket{\psi_{t+1}}$ in the legal subspace, and a vector that lies in the space of locally detectable illegal configurations. For the endpoint states, we only get
$H_{\textrm{prop}} \ket{\psi_0} = \ket{\psi_0} - \ket{\psi_1} + $ illegal states, and $H_{\textrm{prop}} \ket{\psi_K} = \ket{\psi_K} - \ket{\psi_{K-1}} +$ illegal states. Observe that within the legal subspace, both the rows and columns of the matrix form of $H_{\textrm{prop}}$ sum to zero. Looking now at a history state $\ket{\eta}$, a uniform superposition of legal states (for a given initial state), we obtain $\bra{\eta}H_{t}\ket{\eta} = 0$, since $H_{\textrm{prop}}\ket{\eta} = 0$ when restricted to the legal subspace, and the illegal terms produced by $H_{\textrm{prop}}$ are orthogonal to $\ket{\eta}$. $\square$
  
With the propagation Hamiltonian evaluating to zero on
a proper history state, we have 
$$
\langle \eta \vert J_{\mathrm{in}}H_{\mathrm{in}} +
J_{\mathrm{prop}}H_{\mathrm{prop}} + J_{\mathrm{pen}}H_{\mathrm{pen}}
\vert \eta \rangle = 0. 
$$
Consider now the last remaining term in the Hamiltonian, $\langle \eta \vert H_{\mathrm{out}}
\vert \eta \rangle$. Since $H_{\mathrm{out}}$ acts only on the state
with $\gate$ on the last qubit, this term equals $\frac{1}{K+1}
\langle \psi_K \vert H_{\mathrm{out}} \vert \psi_K \rangle
=\frac{p_0}{K+1}$, where $p_0$ is the probability that the output qubit is zero in the final
step. Since the computation accepts with probability $1-p_0 \geq 1 -
\epsilon$, we have $\bra{\eta}H_{\mathrm{out}}\ket{\eta} \leq
  \frac{\epsilon}{K+1} $. Finally,  
\begin{align}
	\bra{\eta} H \ket{\eta} = 
	\bra{\eta}
	 J_{\mathrm{in}}H_{\mathrm{in}} +
	J_{\mathrm{prop}}H_{\mathrm{prop}} + J_{\mathrm{pen}}H_{\mathrm{pen}} + H_{\textrm{out}}
	\ket{\eta}
	\leq \frac{\epsilon}{K+1}.
\end{align} 
This concludes our completeness proof, showing that the energy for a proper history state corresponding
to the verifying computation on a well-accepted witness $\ket{\xi}$ is close to zero. Therefore, the ground state energy of $H$ is also small, in fact it is upper-bounded by $a = \frac{\epsilon}{K+1}$, where $K = (R-1)(3n^2+2n-1)+2n$, with $R$ a polynomial in $n$.


\section{Soundness}
\label{sec:sound}

We now need to show that in the case that there exists no witness that
is accepted by $V_x$ with probability greater than $\epsilon$, the ground state energy of $H$ is bounded away from zero.

We partition the set of configurations into minimal sets $S$ that are
invariant under the action of $H_\mathrm{prop}$. The invariant sets
are of three types: 
\begin{enumerate}
\item Sets that contain legal configurations and locally detectable illegal configurations.
\item Sets that contain only locally detectable illegal configurations.
\item Sets that contain only illegal configurations, some of which are not locally detectable
\end{enumerate}
As we have seen previously, the action of $H_\mathrm{prop}$ on a legal
configuration produces legal `forward' and `backward' transitions,
besides transitions to illegal configurations. 

Illegal configurations that are not locally detectable either have the
wrong number of qubits or have the right number of qubits but are
incorrectly aligned\footnote{An example of a locally undetectable, misaligned sequence: $\mgdots \dead \qubit \bdry \insi \qubit \parity \insi \gate \parity \qubit \blnk
\parity \blnk \blnk \bdry  \blnk \blnk \mgdots$
The block length is 4, and when the $\gate$ particle eventually reaches the front of the [{\em qubits}], it will happen away from a sequence boundary -- there we'll be able to detect it.} with the blocks. Since the transition rules do not
change the number of two-state sites in a configuration nor change the
alignment of the $[qubits]$ string, legal configurations cannot
transition to illegal configurations that are not locally detectable.
Similarly, illegal configurations that are not locally detectable can
only turn into other non-locally detectable illegal configurations, or
into locally detectable illegal configurations.  

A vector belonging to a subspace of type 2 has energy $\geq
J_\mathrm{pen}$, due to the presence of at least one locally
detectable illegal pair that is penalized by $H_\mathrm{pen}$. We now
need to show that vectors from spaces of type 1 and 3 have high
energy. We do this by repeated use of the Projection Lemma, a
technique introduced in~\cite{KKR06}. The lemma allows us to
bound the ground state energy of our Hamiltonian by restricting it to
progressively smaller subspaces of the Hilbert space. 

\vspace*{12pt}
\noindent
\begin{lemma}[Projection Lemma, ~\cite{KKR06} Lemma 1]
  Let $H = H_1 + H_2$ be the sum of two Hamiltonians operating on some
  Hilbert space $\mathcal{H} = S + S^\perp$.  Suppose the Hamiltonian
  $H_2$ is such that $S$ is a zero eigenspace and the eigenvectors in
  $S^\perp$ have eigenvalue at least $J> 2||H_1||$.  Then,
\begin{equation}
\lambda(H_1 \vert_S) - \frac{\Vert H_1 \Vert^2}{J - 2\Vert H_1 \Vert}
\leq \lambda(H) \leq \lambda(H_1\vert_S)
\end{equation}
where $\lambda(A)$ denotes the lowest eigenvalue of an operator $A$.
\end{lemma}

\vspace*{12pt}
\noindent

\subsection{Type 1 subspace}
We consider the action of $H$ on $\mathcal{H}_1$, the space spanned by
configurations of type 1. We apply the projection lemma with 
$$
H_1 = J_{\mathrm{in}}H_{\mathrm{in}} +
J_{\mathrm{prop}}H_{\mathrm{prop}} + H_{\mathrm{out}}, \  H_2 =
J_{\mathrm{pen}}H_{\mathrm{pen}}.
$$
Let $S_{\textrm{pen}}$ be the subspace of $\mathcal{H}_1$ that is spanned by
legal configurations.  Then $S_{\textrm{pen}} \subseteq \mathcal{H}_{1}$ is the
zero eigenspace of $H_2$. On $S_{\textrm{pen}}^{\perp}$, $H_2$ has energy $\geq
J_{\mathrm{pen}}$. Since $\Vert H_1 \Vert \leq poly(n)$, we can pick
$J_{\mathrm{pen}}$ to be some polynomial such that $J_{\mathrm{pen}} >
2 \Vert H_1 \Vert$, allowing us to apply the projection lemma:
\begin{equation} \label{ineq1}
\lambda (H) \geq \lambda (H_{1}\vert_{S_{\textrm{pen}}}) - 1/8.
\end{equation}

Now we bound the lowest eigenvalue of $H_{1}\vert_{S_{\textrm{pen}}}$, 
$$
H_{1}\vert_{S_{\textrm{pen}}} =  H_{\mathrm{out}}\vert_{S_{\textrm{pen}}} +
J_{\mathrm{in}}H_{\mathrm{in}}\vert_{S_{\textrm{pen}}} +
J_{\mathrm{prop}}H_{\mathrm{prop}}\vert_{S_{\textrm{pen}}} .
$$
 We apply the projection lemma again, with 
$$
 H'_{1} = H_{\mathrm{out}}\vert_{S_{\textrm{pen}}}  +
 J_{\mathrm{in}}H_{\mathrm{in}}\vert_{S_{\textrm{pen}}}, \  H'_{2} =
 J_{\mathrm{prop}}H_{\mathrm{prop}}\vert_{S_{\textrm{pen}}}.
$$

 To simplify the analysis of the eigenvalues of $H_{\mathrm{prop}}$,
 we rotate to a different basis -- one in which all the gates from
 $V_x$ are just the identity operator. We define the unitary matrix
 $W$:
$$
W = \sum_{t=0}^{K}U_{t} \cdots U_{1}\otimes \vert t \rangle \langle t \vert
$$
where $\vert t \rangle$ represents the configuration in the $t$-th
step of the computation, and $U_{t},\dots, U_{1}$ are the first $t$
unitary operations performed on the qubit content of the
particles. Then we have,  
\begin{align}
W^{\dagger} H_{\mathrm{prop}}\vert_{S_{\textrm{pen}}} W = I \otimes
\begin{bmatrix}
\frac{1}{2} & -\frac{1}{2} & 0 & 0 &\cdots & 0 & 0 \\[0.3em]
-\frac{1}{2} & 1 & -\frac{1}{2} & 0 & \cdots & \cdots & 0 \\[0.3em]
0 & -\frac{1}{2} & 1 & -\frac{1}{2} & \ddots &  &  \\[0.3em]
0 & 0 & -\frac{1}{2} & \ddots & \ddots &  & \vdots \\[0.3em]
\vdots & & & \ddots & & & 0 \\[0.3em]
0 & & & & & 1 & -\frac{1}{2} \\[0.3em]
0 & 0 &  &\cdots & 0 & -\frac{1}{2} & \frac{1}{2}
\end{bmatrix}_{(K+1)\times(K+1)}. \label{Wmatrix}
\end{align}

This matrix has only one zero-eigenvector, namely the valid history
state.  Therefore $S_{\textrm{prop}} \subset S_{\textrm{pen}} \subset \mathcal{H}_1$ the
set of correct history states (disregarding initial state of ancilla
qubits).  This matrix has second largest eigenvalue $\geq
\frac{1}{2(K+1)^2}$ (see Appendix~\ref{app:eigen}). Therefore, in
$S_{\mathrm{prop}}^{\perp}$, $H'_{2}$ has minimum energy $\geq
J_{\mathrm{prop}}\cdot \frac{1}{2(K+1)^2}$. Choosing
$J_{\mathrm{prop}}$ so that $\frac{J_{\mathrm{prop}}}{2(K+1)^2} >
2\Vert H'_{1} \Vert$, the projection lemma gives us:
\begin{equation} \label{ineq2}
\lambda(H_{1}\vert_{S_{\textrm{pen}}}) \geq \lambda (H'_{1}\vert_{S_{\textrm{prop}}}) - \frac{1}{8}.
\end{equation}

We now apply the projection lemma a third time, to $H'_{1}\vert_{S_{\textrm{prop}}}$, with
$$
H''_{1} = H_{\mathrm{out}}\vert_{S_{\textrm{prop}}}, \  H''_{2} = J_{\mathrm{in}}H_{\mathrm{in}}\vert_{S_{\textrm{prop}}}.
$$
$H''_{2}$ has zero eigenspace $S_{in} \subset S_{\textrm{prop}} \subset S_{\textrm{pen}}
\subset \mathcal{H}_1$, the set of history states with correctly
initialized ancilla qubits. Since $H''_2$ is in the standard basis and
applies to vectors that are history states with 0 on the ancilla input
(i.e., in $S_{in}^{\perp}$), $H''_{2}$ has minimum energy
$O\left(\frac{1}{K+1}\right)$. Therefore, $J_{\mathrm{in}}$ can be
chosen so that $\frac{J_{\mathrm{in}}}{K+1} > 2\Vert H''_{1}
\Vert$. Then
\begin{equation} \label{ineq3}
\lambda(H'_{1}\vert_{S_{\textrm{prop}}}) \geq \lambda (H''_{1}\vert_{S_{in}}) - \frac{1}{8} ,
\end{equation}
and $H''_{1}\vert_{S_{in}} = H_{\mathrm{out}}\vert_{S_{in}}$. For any
input state $|\xi,0\rangle$, the circuit $V_{x}$ accepts with
probability $< \epsilon$. Therefore, for any $|\eta\rangle \in
S_{in}$, we have $\langle \eta \vert H_{\mathrm{out}} \vert \eta
\rangle > (1- \epsilon)/(K+1)$.  In particular, this is true for the
eigenvector $|\eta\rangle$ of $H_{\mathrm{out}}$ with the lowest
eigenvalue.  Therefore $\lambda(H_{\mathrm{out}}|_{S_in}) \geq
(1-\epsilon)/(K+1)$.

Combining \eqref{ineq3} with \eqref{ineq1} and \eqref{ineq2}, we have
$
\lambda(H) \geq \frac{5}{8} - \epsilon.
$

Now we look at vectors from subspaces of type 3. 


\subsection{Type 3 subspace}
\label{locallyundetectablesection}

A locally undetectable illegal configuration either $(i)$ has the wrong number of qubits, or $(ii)$ has the $[qubits]$ string incorrectly aligned with the blocks. 


Consider a (locally undetectable illegal) configuration with a $\gate$ site. The $[qubits]$ string either crosses a block boundary, or is too short (or both). The $\gate$ moves right until it either hits the end of the $[qubits]$ string or it hits a block boundary. If the end of the $[qubits]$ string does not coincide with a block boundary, the configuration eventually evolves to contain either $\parity \gate \insi \bdry$ or $\parity \gate \blnk \parity$, both of which are  locally detectable. If the end of the $[qubits]$ string does coincide with the block boundary (this can only happen when the $[qubits]$ string is too short), the qubits get moved over into the next block, where a $\gate$ is generated again, and, moving right, eventually produces a pair $\parity \gate \blnk \parity$, which is locally penalized. 

A locally undetectable illegal configuration with a $\lmove$ also eventually evolves into a locally detectable one: the $\lmove$ moves the $[qubits]$ string to the right until the beginning of the $[qubits]$ string coincides with the beginning of a block, and generates a $\gate$, at which point the above argument applies. 

If our locally undetectable illegal configuration has neither $\gate$ or $\lmove$, i.e. its [{\em qubits}] substring consists only of $\qubit$s (separated by $\insi$s), the qubits begin to move themselves to the right, eventually generating a $\lmove$ or $\gate$ flag, at which point, the previous arguments apply: the evolution does indeed result in a locally detectable configuration. 

In all the above cases, a locally detectable illegal configuration is reached within polynomially many steps/transitions. To see this, consider a configuration with $n'$ qubits. It takes $poly(n')$ steps to move the $[qubits]$ string over one block, and by the preceding arguments, a locally checkable configuration must be reached at some point in this `round' of computation. Since $n'$ can be at most $2nR$, this number of steps (which we label $K'$) must be polynomial in $n$. In other words, the transition rules eventually take the state outside $\mathcal{H}_3 = $ Span(configurations of type 3). We can treat the restriction of $H_{\mathrm{prop}}$ to $\mathcal{H}_3$ in much the same way as we did its restriction to $\mathcal{H}_1$. 



We attempt to bound the lowest eigenvalue of
$J_{\mathrm{pen}}H_{\mathrm{pen}} +
J_{\mathrm{prop}}H_{\mathrm{prop}}$ on $\mathcal{H}_3
$  using the projection lemma,
with $H_{1} = J_{\mathrm{prop}}H_{\mathrm{prop}}$ and $H_{2} =
J_{\mathrm{pen}}H_{\mathrm{pen}}$. The zero eigenspace of $H_{2}$ is
the space of illegal states that are \emph{not} locally detectable,
$S_{\textrm{pen}} \subset \mathcal{H}_3$. $H_{2}$ has energy $\geq
J_{\mathrm{pen}}$ on $S_{\textrm{pen}}^{\perp}$. Choosing $J_{\mathrm{pen}}$ to
be $poly(J_{\mathrm{prop}}\Vert H_{\mathrm{prop}}\Vert)$,  
\begin{equation}
\lambda(H) \geq \lambda(H_{1}\vert_{S_{\textrm{pen}}}) - \frac{1}{8}.
\end{equation}
We now need the lowest eigenvalue of
$J_{\mathrm{prop}}H_{\mathrm{prop}}\vert_{S_{\textrm{pen}}}$. We rotate bases
once again, using the unitary matrix $W$ defined earlier, with the
difference that $\vert t \rangle$ now represents the $t$-th
configuration in the sequence of (locally undetectable) illegal
configurations that arises from the transition rules and forms the
steps of the `computation'. This sequence of configurations terminates
in a locally detectable illegal configuration on at least one
end. (The other end of the sequence could be a locally undetectable
illegal configuration from which there are no further transitions.)
When we have a transition to a locally illegal configuration, the
action of $H_{\mathrm{prop}}$ on the last nonlocally detectable
illegal configuration ($\vert K'\rangle$) in the sequence is to pick
this configuration out twice, i.e., there are two projectors onto this
configuration in $H_{\mathrm{prop}}\vert_{S_{\textrm{pen}}}$, with the result
that, in the space of configurations, the last (or first, or both) diagonal
element of $W^{\dagger}H_{\mathrm{prop}}\vert_{S_{\textrm{pen}}}W$ is 1 instead of $\frac{1}{2}$.  

In other words, $W^{\dagger}H_{\mathrm{prop}}\vert_{S_{\textrm{pen}}}W$ looks like 
\begin{equation}
 I \otimes
\begin{bmatrix}
f & -\frac{1}{2} & 0 & 0 &\cdots & 0 & 0 \\[0.3em]
-\frac{1}{2} & 1 & -\frac{1}{2} & 0 & \cdots & \cdots & 0 \\[0.3em]
0 & -\frac{1}{2} & 1 & -\frac{1}{2} & \ddots &  &  \\[0.3em]
0 & 0 & -\frac{1}{2} & \ddots & \ddots &  & \vdots \\[0.3em]
\vdots & & & \ddots & & & 0 \\[0.3em]
0 & & & & & 1 & -\frac{1}{2} \\[0.3em]
0 & 0 &  &\cdots & 0 & -\frac{1}{2} & g
\end{bmatrix}_{(K'+1)\times(K'+1)}
\end{equation}
with $f=1, g=\frac{1}{2}$ or $f=g=1$. This is a matrix for a quantum walk on a line with particular boundary conditions.

The least eigenvalue of either of these matrices is
$O\left(\frac{1}{(K'+1)^2} \right)$ (see Appendix~\ref{app:eigen}). Since $K'$ is a polynomial in
$n$,   choosing $J_{\mathrm{prop}}$ to be an appropriately large
polynomial in $n$, we can lower-bound the energy of $H$ on type-3
subspaces by some constant.  


\bibliographystyle{plain}
\bibliography{localham}


\appendix{~~Eigenvalues}
\label{app:eigen}

Here we analyze the eigenvalues of the three matrices in Section~\ref{sec:sound}. Our matrices are of the form:
\begin{equation}
\begin{bmatrix}
f & -\frac{1}{2} & 0 & 0 &\cdots & 0 & 0 \\[0.3em]
-\frac{1}{2} & 1 & -\frac{1}{2} & 0 & \cdots & \cdots & 0 \\[0.3em]
0 & -\frac{1}{2} & 1 & -\frac{1}{2} & \ddots &  &  \\[0.3em]
0 & 0 & -\frac{1}{2} & \ddots & \ddots &  & \vdots \\[0.3em]
\vdots & & & \ddots & & & 0 \\[0.3em]
0 & & & & & 1 & -\frac{1}{2} \\[0.3em]
0 & 0 &  &\cdots & 0 & -\frac{1}{2} & g
\end{bmatrix}_{(L+1)\times(L+1)}
\end{equation}
where, in subspaces of type 1, $f = g = \frac{1}{2}$, and in subspaces of type 3, either (i) $f = g = 1$ or (ii) $f = 1$ and $g = \frac{1}{2}$ (we could also have $f = \frac{1}{2}$ and $g = 1$, but this doesn't change the eigenvalues).

We wish to solve the eigenvalue equation $Mx = \lambda x$, where $x = (x_0, x_1, x_2, \ldots, x_L)^{T}$ and $M$ is our matrix. It is easy to see that $x$ must satisfy the equations
\begin{equation}
-\frac{1}{2} x_{j-1} + x_j -\frac{1}{2} x_{j+1} = \lambda x_{j} \qquad \text{for } 1 \leq j \leq L-1 .
\label{nonbc}
\end{equation}
We use the ansatz 
\begin{equation}
 x_j = A \cos k(j+c) + B \sin k(j+c). 
\end{equation}
where $A$, $B$, $k$ and $c$ are reals. Plugging this into~\eqref{nonbc}:
\begin{align*}
-(\cos k) \left( A \cos k(j+c) +B \sin k(j+c) \right) &= (\lambda -1) \left( A \cos k(j+c) + B \sin k(j+c) \right) \\
\lambda &= 1 - \cos k 
\end{align*}

Any vector $x$ with $x_j = A \cos k(j+c) + B \sin k(j+c)$ satisfies \eqref{nonbc} with $\lambda = 1 - \cos k$. 
Now we apply the `boundary conditions' in the first case ($f = g =\frac{1}{2}$). The eigenvectors and eigenvalues are 
\begin{align*}
\lambda_m &= 1 - \cos\left( \frac{m\pi}{L+1} \right), \\
\ket{\psi_m} &= \sum_{j=0}^{L} \cos \left( \frac{m \pi}{L+1} (j+\frac{1}{2}) \right) \ket{j} \qquad \text{for } m = 0,1,\ldots, L.
\end{align*}
We can check this by plugging into the boundary condition equations 
\begin{align*}
\frac{1}{2} x_0 - \frac{1}{2} x_1 &= (1 - \cos k) x_0, \\
- \frac{1}{2} x_{L-1} + \frac{1}{2} x_L &= (1 - \cos k) x_L. 
\end{align*}
The lowest eigenvalue in this case is 0 (when $m = 0$). The second-lowest eigenvalue is 
\begin{align*}
1 - \cos \left( \frac{\pi}{L+1} \right) &> \left( \frac{1}{L+1} \right)^2 \left(  \frac{\pi^2}{2!} -\frac{\pi^4}{4!(L+1)^2} \right) \\
& = \Omega\left(  \frac{1}{(L+1)^2} \right)
\end{align*}
i.e., $1/poly(L)$, where $L$ is the number of steps, as promised. 

Now we consider the next set of boundary conditions $f = g =1$ (subspace of type 3). We get eigenvalues and eigenvectors:
\begin{align*}
\lambda_m &= 1 - \cos\left( \frac{(m+1)\pi}{L+1} \right), \\
\ket{\psi_m} &= \sum_{j=0}^{L} \cos \left( \frac{(m+1) \pi}{L+2} (j+1) \right) \ket{j} \qquad \text{for } m = 0,1,\ldots, L.
\end{align*}
This is again easily checked by plugging into the equations
\begin{align*}
 x_0 - \frac{1}{2} x_1 &= (1 - \cos k) x_0, \\
- \frac{1}{2} x_{L-1} + x_L &= (1 - \cos k) x_L. 
\end{align*}
The lowest eigenvalue here is $\lambda_0 = 1 - \cos \left( \frac{\pi}{L+2} \right)$, which is $\Omega (\frac{1}{(L+2)^2})$. 

The final set of boundary conditions to consider is $f = 1, g=\frac{1}{2}$. The eigenvalues and eigenvectors in this case are:
\begin{align*}
\lambda_m &= 1 - \cos\left( \frac{(2m+1)\pi}{2L+3} \right), \\
\ket{\psi_m} &= \sum_{j=0}^{L} \sin \left( \frac{(2m+1) \pi}{2L+3} (j+1) \right) \ket{j} \qquad \text{for } m = 0,1,\ldots, L.
\end{align*}
The lowest eigenvalue here is $\lambda_0 = 1 - \cos \left( \frac{\pi}{2L+3} \right) = \Omega(\frac{1}{(2L+3)^2})$.

\end{document}